\documentclass[letterpaper,twocolumn,10pt]{article}
\usepackage{zhanggroup}

\usepackage[absolute]{textpos}
\usepackage{caption}
\usepackage{subcaption}
\usepackage{float}
\usepackage{authblk}
\usepackage{placeins}
\usepackage{multirow}
\usepackage{booktabs} 
\usepackage{algorithmic}
\usepackage{amsmath,amsfonts}
\usepackage{textcomp}
\usepackage{xcolor}
\usepackage{xspace}
\usepackage{xurl}
\usepackage{amsmath}
\usepackage{tikz}
\usepackage{filecontents}
\usepackage{booktabs} 
\newcommand{\MIA}{\mathcal{I}}
\newcommand{\MLModel}{\mathcal{M}}

\newcommand{\AttackModel}{\mathcal{A}}

\newcommand{\SeqMIA}{\texttt{SeqMIA}}
\newcommand{\mypara}[1]{\noindent\textbf{#1.}\xspace}

\newcommand{\tablescale}{0.8}

\begin{document}

\begin{textblock}{15}(1.9,1)
To Appear in 2024 ACM SIGSAC Conference on Computer and Communications Security, October 14-18, 2024
\end{textblock}
\date{}

\title{\bf SeqMIA: Sequential-Metric Based Membership Inference Attack}

\author[1]{Hao Li\thanks{The first two authors made equal contributions.}}
\author[2]{Zheng Li\thanksmark{1}}
\author[1]{Siyuan Wu}
\author[1]{Chengrui Hu}
\author[1,3]{Yutong Ye}
\author[1]{Min Zhang\thanks{Corresponding author.}}
\author[1]{Dengguo Feng}
\author[2]{Yang Zhang}

\affil[1]{Trusted Computing and Information Assurance Laboratory, Institute of Software, Chinese Academy of Sciences}
\affil[2]{CISPA Helmholtz Center for Information Security}
\affil[3]{Zhongguancun Laboratory}

\maketitle

\begin{abstract}
Most existing membership inference attacks (MIAs) utilize metrics (e.g., loss) calculated on the model's final state, while recent advanced attacks leverage metrics computed at various stages, including both intermediate and final stages, throughout the model training.
Nevertheless, these attacks often process multiple intermediate states of the metric independently, ignoring their time-dependent patterns. 
Consequently, they struggle to effectively distinguish between members and non-members who exhibit similar metric values, particularly resulting in a high false-positive rate.

In this study, we delve deeper into the new membership signals in the black-box scenario. 
We identify a new, more integrated membership signal: \textit{the Pattern of Metric Sequence}, derived from the various stages of model training.
We contend that current signals provide only partial perspectives of this new signal: the new one encompasses both the model's multiple intermediate and final states, with a greater emphasis on temporal patterns among them.
Building upon this signal, we introduce a novel attack method called Sequential-metric based Membership Inference Attack (\SeqMIA). 
Specifically, we utilize knowledge distillation to obtain a set of distilled models representing various stages of the target model's training. 
We then assess multiple metrics on these distilled models in chronological order, creating \textit{distilled metric sequence}. 
We finally integrate distilled multi-metric sequences as a sequential multiformat and employ an attention-based RNN attack model for inference.
Empirical results show \SeqMIA\ outperforms all baselines, especially can achieve an order of magnitude improvement in terms of TPR @ 0.1\% FPR.
Furthermore, we delve into the reasons why this signal contributes to \SeqMIA's high attack performance, and assess various defense mechanisms against \SeqMIA.\footnote{Our code is available at \url{https://github.com/AIPAG/SeqMIA}}
\end{abstract}

\section{Introduction}

Machine learning (ML) has developed rapidly in the past decade. Unfortunately, existing studies~\cite{fredrikson2015model, ganju2018property,shokri2017membership} have shown that ML models can leak private information about their training set. 
Membership inference attacks (MIAs)~\cite{shokri2017membership} is one of the main privacy attacks that have attracted lots of researchers' concerns. 
It aims to infer whether a sample belongs to a model's training set, which in turn violates the privacy of the sample's owner. 
For example, if an ML model is trained on data collected from individuals with a certain disease, an adversary who knows that a victim's data belongs to the training data of the model can quickly infer the victim's health status.

\newcommand{\size}{1.65in}
\begin{figure}[t]
  \centering
  \begin{subfigure}[t]{\size}
    \centering
    \includegraphics[width=\size]{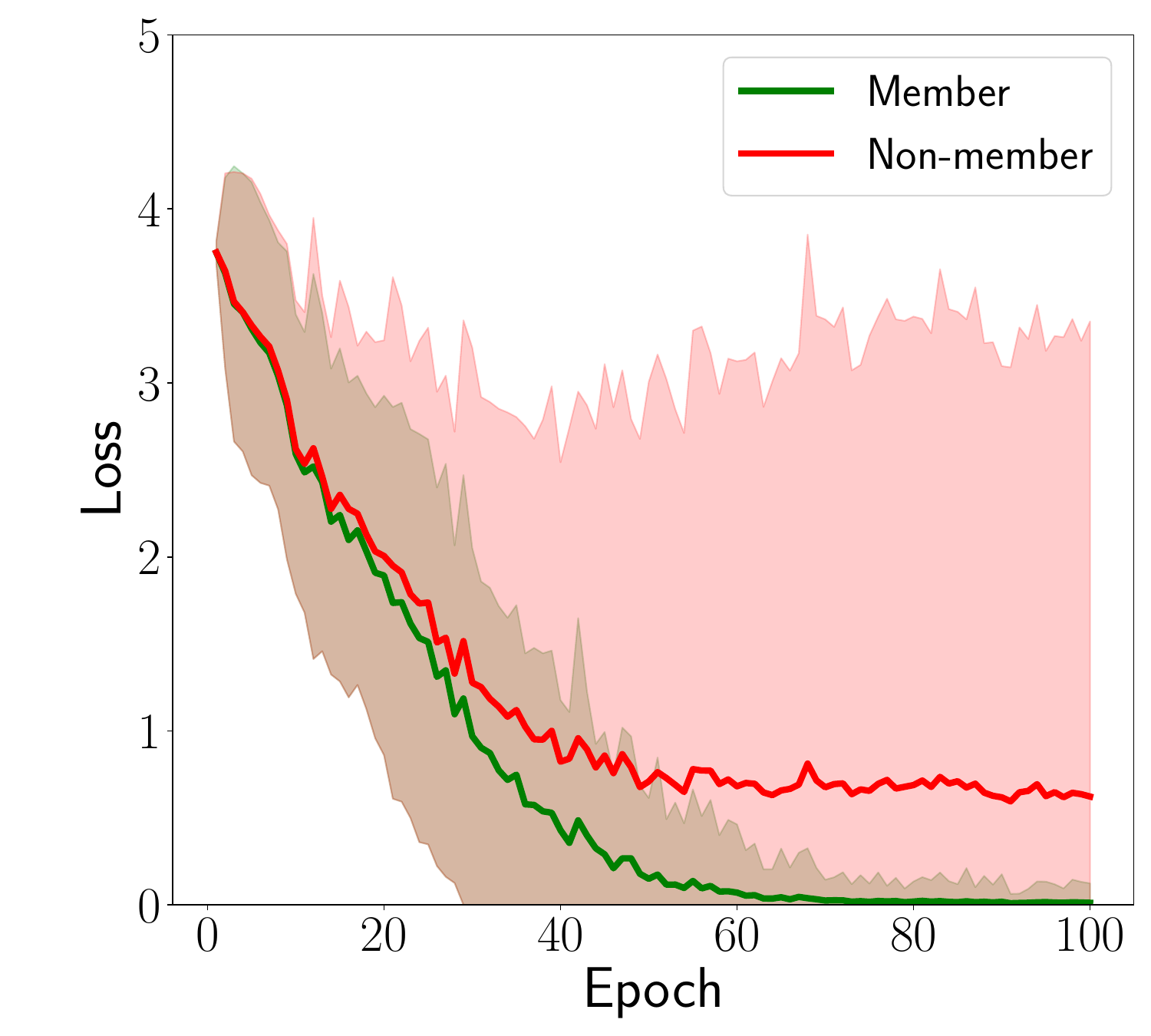}
    \caption{Fluctuation Area of Loss}
    \label{FluctuationofLoss}
  \end{subfigure}
  \begin{subfigure}[t]{\size}
    \centering
    \includegraphics[width=\size]{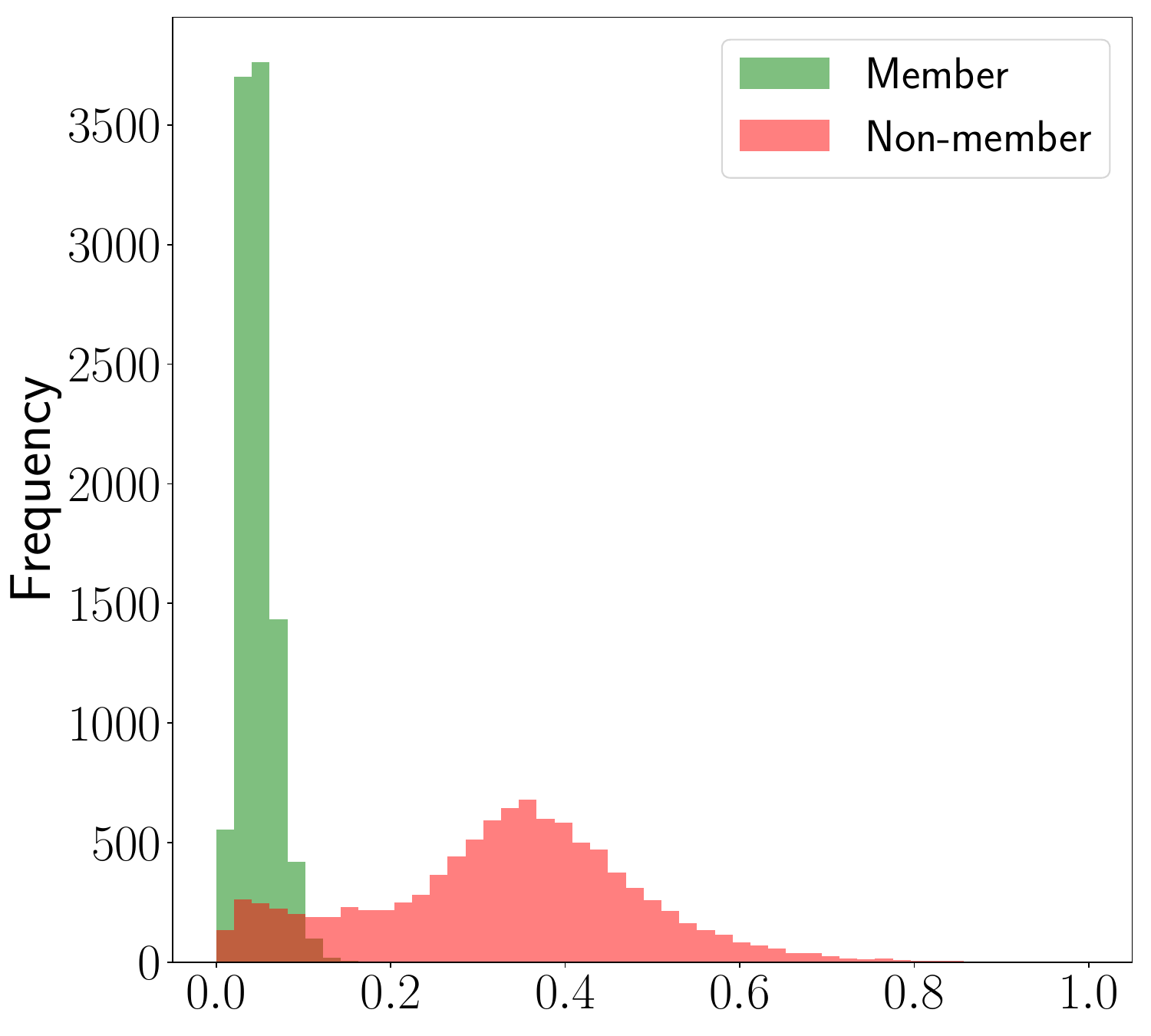}
    \caption{Distribution of CLFA}
    \label{FluctuationDistribution}
  \end{subfigure}
  \caption{(a) the mean curves and fluctuation area of loss values for members and non-members during different training epochs; (b) the distribution of the cumulative loss fluctuation amplitude (CLFA) within 100 epochs.
  }
  \label{metric-curves-loss-entropy}
\end{figure}

\begin{table*}[t]
    \caption{Various perspectives of the membership signal ``Pattern of Metric Sequences.''
    ``\checkmark'' means this attack is based on this perspective and ``-'' indicates that it is not.}
    \centering
    \scalebox{0.76}{
        \begin{tabular}{l|c|c|c|c|c|c}
        \toprule
        \multirow{4}{*}{Attacks} & \multicolumn{6}{c}{Pattern of Metric Sequences (i.e., metric values overall training epochs)}\\
        \cmidrule(r){2-7}
         & \multirow{2}{*}{Final State}  & \multirow{2}{*}{Middle States}  &  \multicolumn{4}{c}{Time-Dependent Patterns} \\
         \cmidrule(r){4-7}
         & &  & Fluctuation  & Correlation  & Decline Rate & Other Possible Implicit Patterns \\
        \midrule
        ~\cite{shokri2017membership, salem2018ml,song2021systematic, sablayrolles2019white, watson2021importance, carlini2022membership, ye2022enhanced, liu2022membership, bertran2024scalable} & \checkmark & -- & --  & -- & -- & -- \\
        \midrule
        TrajectoryMIA~\cite{liu2022membership} & \checkmark & \checkmark  & -- & -- & -- & --  \\
        \midrule
        \SeqMIA & \checkmark & \checkmark & \checkmark & \checkmark & \checkmark & \checkmark \\
        \bottomrule
        \end{tabular}
    }
    \label{tab:signals}
\end{table*}

Most existing studies~\cite{shokri2017membership, salem2018ml, song2021systematic, zhang2021membership, hisamoto2020membership, chen2021machine} employ the target model's output posteriors or some metric (e.g., loss) derived from them to launch their attacks. 
These attacks demonstrate effectiveness in average-case metrics such as balanced accuracy and ROC-AUC due to members typically exhibiting smaller losses compared to non-members. 
However, these attacks exhibit a high false-positive rate (FPR) when encountering both members and non-members with similar small losses.
A high false positive rate means that an attacker will incorrectly identify non-member samples as members, thereby reducing the attack's effectiveness and reliability.

To tackle this issue, recent studies~\cite{carlini2022membership, ye2022enhanced} have employed sample-dependent thresholds to calibrate membership inference based on the target model, i.e., final model state at 100th (see \autoref{FluctuationofLoss}).
An alternative approach, known as TrajectoryMIA~\cite{liu2022membership}, introduces an additional membership signal, which is a collection of loss values gathered during the target model training process (i.e., 0$\sim$100th epochs). 
The loss value set derived from various model states can reveal greater distinctions between members and non-members, even when they show similar low losses in the final model state.
The findings of our experiments, however, indicate that these recent studies still face challenges in effectively distinguishing between members and non-members with similar sets of loss values, leading in particular to significantly higher false positive rates (FPR).

\subsection{Our contributions}
To overcome these limitations, in this work, we make an attempt to answer, ``is it possible to explore a new membership signal that enhances the distinguishability between members and non-members, with a specific focus on reducing false-positive rate?''

Fortunately, we have discovered a new membership signal termed \textit{the Pattern of Metric Sequence}, which is also derived from the various stages of model training. 
As \autoref{tab:signals} illustrates, we claim that the aforementioned signals provide only partial perspectives of this new signal: this new signal includes both the model's multiple intermediate and final states and focuses more on time-dependent patterns among them.
To our knowledge, this signal has not been previously recognized or utilized in prior literature. 
Intuitively, we verify this signal from time-dependent views, such as fluctuation, correlation, and decline rate. 
We now illustrate the first two perspectives (see decline rate in \autoref{addtional-patterns}).

\mypara{Fluctuation of Metric Sequences} We choose the most commonly used metric, loss, as our example.
\autoref{FluctuationofLoss} shows the sequence of loss values as the training progresses (denoted as loss sequence).
Interestingly, we have further observed a new difference: \textit{the fluctuation of loss sequence} between members and non-members also exhibits significant differences. 
More concretely, the loss sequence fluctuation of members tends to be smaller than that of non-members, especially around the 60th to 100th epoch.
Besides expressing such fluctuation qualitatively, we further measure them quantitatively.
Specifically, we compute the cumulative loss fluctuation amplitude (CLFA) for each sample by measuring the loss variation across \textit{consecutive epochs}. 
We then count the frequency of the samples regarding their CLFA distribution.
As depicted by \autoref{FluctuationDistribution}, we observe members exhibit significantly smaller fluctuation of loss sequence compared to non-members.
The results confirm that there exists a very clear difference in the pattern of loss sequence between members and non-members (see other metrics in Appendix \autoref{appendix-metric-fluctuation}).
Note that this observation is time-dependent and can only be observed in metric sequence as the training epoch progresses, unlike the \textit{loss set} used in TrajectoryMIA, where shuffling the order does not affect the attack performance (see \autoref{Analysis}). 

\begin{figure}[t]
  \centering
  \includegraphics[width=0.96\linewidth]{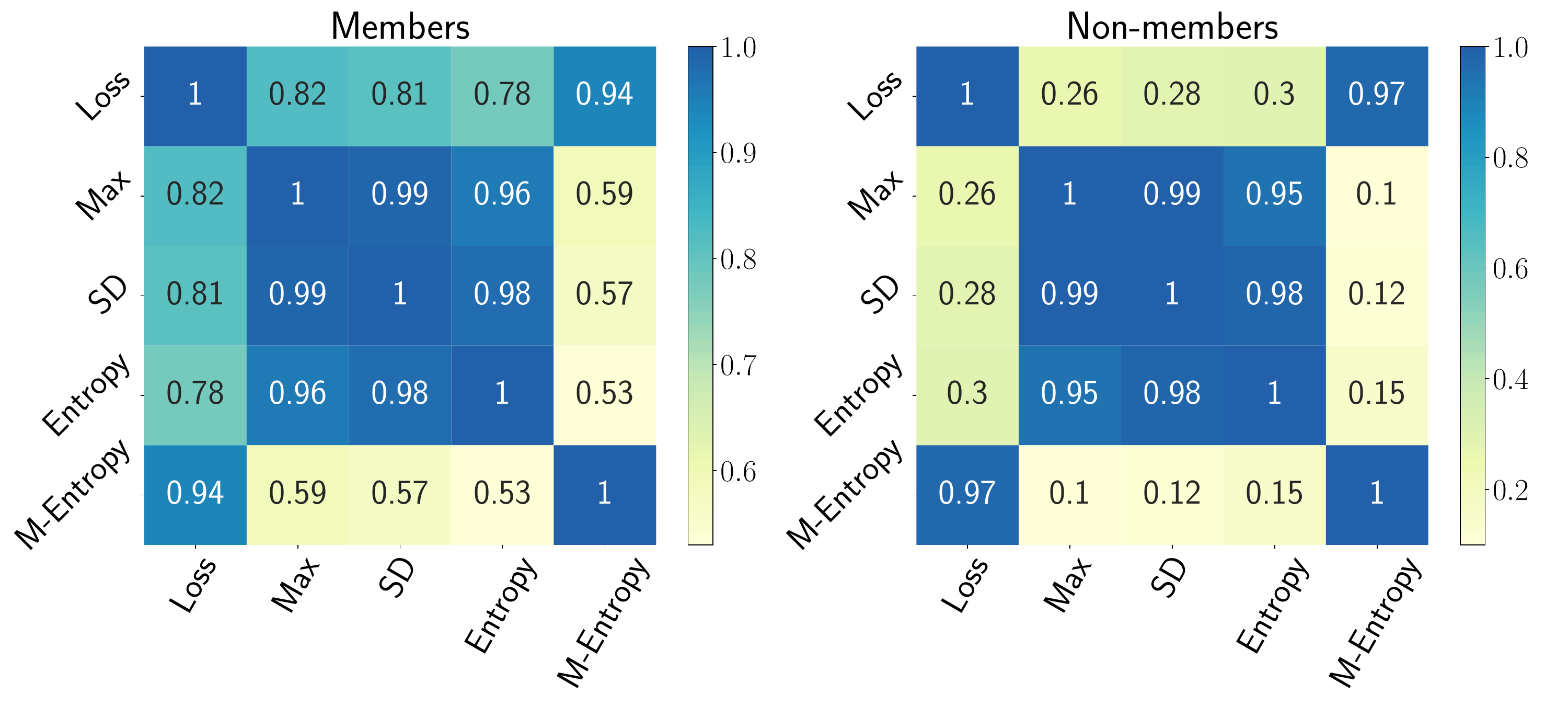}
  \caption{Absolute correlation coefficients among multiple metrics calculated from MLPs trained on Location.}
  \label{metric-correlations}
\end{figure}

\mypara{Correlation between Metric Sequences}
Building upon the metric sequence, we delve into another new view: the correlation between two different metric sequences.
The intuition is that metric sequences of members tend to follow a similar trend compared to non-members, as the model is trained on members. 
\autoref{metric-correlations} presents the correlation coefficients among multiple sequences metrics. 
We observe that every pair of metric sequences for members shows correlation coefficients no smaller than those for non-members and, in most cases, even larger ones.

\mypara{\SeqMIA} Building upon the pattern of metric sequences, we introduce a novel membership inference attack named \SeqMIA\ (Sequential-metric based Membership Inference Attack). 
First, the adversary employs knowledge distillation to obtain a set of distilled models representing various stages of the target model's training. Then, the adversary assesses multiple metrics on these distilled models in chronological order, creating \textit{distilled metric sequence}. 
The adversary integrates multiple distilled metric sequences into a sequential multiformat and utilizes a simple yet highly effective approach to handle this sequential data, namely employing an attention-based recurrent neural network (attention-based RNN) as the attack model for inference.
This attention-based RNN can automatically excavate the aforementioned different patterns of metric sequence (even some more complex implicit patterns) without explicitly characterizing them in advance.

We conduct extensive experiments on 4 popular models using 7 benchmark datasets (4 image and 3 non-image datasets).
Empirical results show that \SeqMIA\ outperforms the baselines in nearly all cases. 
For example, when focusing on VGG-16 trained on CIFAR100, \SeqMIA\ surpasses all baselines by more than an order of magnitude in terms of TPR @ 0.1\% FPR. 
In addition, 
we conduct in-depth comparative analyses of metric non-sequences vs. metric sequences, and single vs. multiple metrics, revealing the reasons for the superior performance of \SeqMIA.
We also conduct ablation studies to analyze various factors on the attack performance.
Finally, we demonstrate that \SeqMIA\ performs better against several defenses compared to the baselines, especially at TPR @ 0.1\% FPR. 
In general, our contributions can be summarized as follows:
\begin{itemize}
    \item We introduce a novel membership signal termed \textit{the Pattern of Metric Sequence}, which can more effectively capture the differences between members and non-members.    
    \item We propose the Sequential-metric based Membership Inference Attack (called \SeqMIA), which acquires sequential multi-metric from the target model's training process using knowledge distillation, then captures the membership signal via attention-based RNN attack model automatically.
    \item We extensively experiment and demonstrate that \SeqMIA\ consistently outperforms all baselines, particularly in reducing the FPR by more than an order of magnitude.
    \item We conduct comprehensive analyses of the features of sequential membership signals, some key factors influencing attack performance, and various defenses against \SeqMIA.
\end{itemize}

\section{Preliminaries}

\subsection{Membership Inference Attack}
Membership inference attack is one of the most popular privacy attacks against ML models.
The goal of the membership inference attack is to determine whether a data sample is used to train a target model.
We consider data samples as members if they are used to train the target model, otherwise, non-members.
Formally, considering a data sample $x$, a trained ML model $\MLModel$, and background knowledge of an adversary, denoted by $\MIA$, the membership inference attack $\AttackModel$ can be defined as the following:
\[
\AttackModel: x, \MLModel, \MIA \rightarrow \{0, 1\}.
\]
Here, 0 means the data sample $x$ is not a member of $\MLModel$'s training dataset, and 1 otherwise. 
The attack model $\AttackModel$ is essentially a binary classifier.

\subsection{Metrics for MIA}\label{metrics}
The success of existing membership inference is attributed to the inherent overfitting properties of ML models, i.e., models are more confident when faced with the data samples on which they are trained.
This confidence is reflected in the model's output posterior, which results in several metrics that effectively differentiate between members and non-members.
We are here to present a brief introduction below:

\mypara{Loss}
Loss, also known as the cost or objective function, measures how well an ML model's predictions match the ground truth for given data samples.  
The goal of the ML algorithm is to minimize this loss, as a lower loss indicates better performance of the model.
Typically, members' losses are much lower than non-members' losses, and most existing works~\cite{yeom2018privacy, sablayrolles2019white, watson2021importance, carlini2022membership, ye2022enhanced} leverage this discrepancy to mount their membership inferences. 

Furthermore, loss trajectory is proposed by ~\cite{liu2022membership}, which is a set of multiple losses from a model's training process, and is implemented as a vector. 
We here emphasize that the loss trajectory is not a sequential signal, due to the fact that there is no order between these loss values. 
If we swap the positions of losses in this vector, we will find that the attack performance of ~\cite{liu2022membership} is unaffected (see \autoref{Analysis}). 
Thus, we denoted it as \textit{loss set} in the following sections in order to clearly indicate its essential features.

\mypara{Max} Max refers to the maximum in the model's output posteriors, which is usually represented as a set of probabilities.
To obtain single predicted class labels from these probabilities, one common approach is to take the class with the highest probability, i.e., maximum value.
Similarly, the maximum value of members is usually greater than that of non-members, which has been used in~\cite{salem2018ml, song2021systematic}.

\mypara{SD} Standard deviation is a measure of the dispersion of the model's output posterior from its mean. 
Members tend to have larger standard deviations than non-members because the model has more confidence in the predictions of the members, i.e., the probability of the correct class is greater, and the probability of the other class is much less.
This metric has been used in~\cite{salem2018ml}.

\mypara{Entropy} Entropy measures the uncertainty or randomness in a model's prediction.
A low entropy indicates that the probability distribution is concentrated and the model is more certain about its predictions, while a high entropy indicates more uncertainty.
Similarly, the entropy value of members is lower than that of non-members, which has been used in~\cite{shokri2017membership, salem2018ml, song2021systematic}

\mypara{M-Entropy} In contrast to entropy, which contains only information about the output posterior, modified entropy (M-Entropy) measures the model prediction uncertainty
given the ground truth label.
Thus, correct prediction with probability 1 leads to a modified entropy of 0, while incorrect prediction with probability 1 leads to a modified entropy of infinity.
Also, the modified entropy of members is usually lower than that of non-members, and this metric is used in~\cite{song2021systematic}.

\subsection{Knowledge Distillation}
Knowledge distillation (KD) is a category of methods that transfer knowledge from large, complex models to smaller, more lightweight ones. 
The primary goal is to improve the performance of the smaller model while reducing resource consumption during deployment.
The main idea is to use the soft information (i.e., the output posterior) of a larger teacher model as a supervised signal to train a smaller student model.
This soft information contains more valuable knowledge than hard ground truth labels, leading to better generalization and efficiency of the student model.

Similar to~\cite{liu2022membership}, we use knowledge distillation to train a distilled model (student model) that is as close as possible to the target model (teacher model). 
In this work, we adopt the most widely-used KD framework proposed by Hinton et al.~\cite{hinton2015distilling}.
Concretely, we use a set of data (called distillation dataset) to query the teacher model and obtain its output posteriors, called soft labels.
Then, when training the student model, soft labels are used to calculate the loss function in addition to the ground truth labels. 
The loss function can be expressed as follows:
\begin{equation}
L = \alpha L_{soft} + (1-\alpha)L_{ground} 
\end{equation}
where $L_{soft}$ is the Kullback-Leibler divergence loss between the soft labels and the student model's output posteriors, $L_{ground}$ is the cross-entropy loss between the student model's output posteriors and the ground truth labels, and $\alpha$ is a weight coefficient.

Note that our goal is to simulate the target model training process and snapshot its intermediate version, rather than transferring knowledge from the larger model to the smaller one.
Therefore, we employ the same model architecture as the target model to build the distilled model.
Further, we set $\alpha = 1$, which means that the distilled model only mimics the target model's output posteriors regardless of the ground truth labels. For the sake of description, the intermediate distilled models obtained by distillation, are named as \textit{snapshots} in this paper.

\section{Attack Methodology}

In this section, we present the attack methodology of \SeqMIA.
We start by introducing the threat model.
Then, we describe the design intuition.
Lastly, we present the detailed pipeline of \SeqMIA.

\subsection{Threat Model}\label{threatmodel}
In this paper, we focus on membership inference attacks in black-box scenarios, which means that the adversary can only access the output of the target model. 
Specifically, we only consider the case where the output is the predicted probability (posterior) rather than the predicted class label. 
Furthermore, we make two assumptions about the adversary's knowledge. 
First, the adversary holds a dataset $D^a$, which is from the same distribution as the target model's training dataset.
Second, the adversary knows the architecture and hyperparameters of the target model. 
Such settings are following previous MIAs~\cite{shokri2017membership, salem2018ml, song2021systematic, yeom2018privacy, long2020pragmatic, carlini2022membership, ye2022enhanced, liu2022membership}. 
Moreover, we further demonstrate in \autoref{ablationstudy} that both of these assumptions can be relaxed.

\subsection{Design Intuition}
As aforementioned, we introduce a new membership signal termed \textit{the Pattern of Metric Sequence}, which is also derived from the various stages of model training. 
This new signal includes both the model's multiple intermediate and final states but focuses more on time-dependent patterns among them.
For example, members' metric sequences tend to demonstrate relatively smaller fluctuations compared to non-members.
In addition, the correlations between different metric sequences of members are also much higher compared to non-members.
Therefore, our general hypothesis is that simultaneous utilization of multiple metric sequences (serialized metric values) would yield significantly stronger membership signals compared to relying solely on a single metric or a non-serialized metric. 
Based on this insight, our first attack strategy is to construct ``multi-metric sequences,'' which carry the pattern of metric sequences. 

Furthermore, the previous study by Liu et al.~\cite{liu2022membership} treats multiple losses from the various model states as a one-dimensional vector and directly feeds it into an MLP attack model for inference. 
However, the MLP model is primarily designed for independent input values and fails to capture the sequential or time-series information present in the input vector. 
This means that the MLP model may overlook important sequence-based signals in the input  (see shuffling the vector's loss values in \autoref{Analysis}).
In contrast, models specifically designed for time-series data, such as Recurrent Neural Networks (RNNs), are better able to capture the sequential information in the input vector, and thus can potentially excavate the sequence-based signals, e.g., fluctuations in the metric sequences as training progresses.
Therefore, our second attack strategy involves using an attention-based RNN as the attack model to process the multiple metric sequences. 
This way, we can automatically uncover not only these explicit patterns but also more complex implicit patterns (see \autoref{Analysis}).

\begin{figure}[t]
  \centering
  \includegraphics[width=\linewidth]{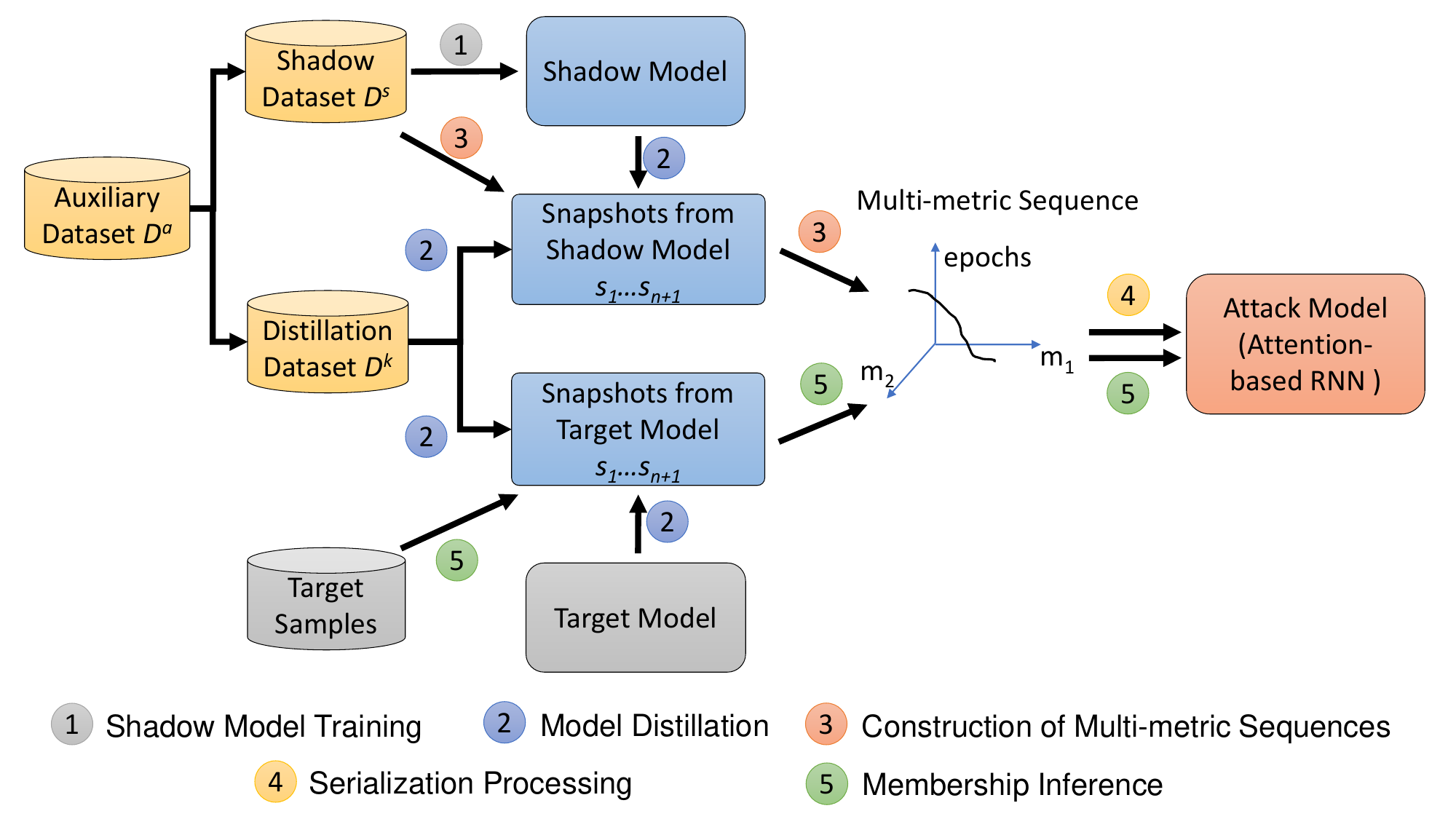}
  \caption{Overview of SeqMIA. Different from existing MIAs, SeqMIA focuses on the sequential membership information (multi-metric sequences) in a high-dimensional space. }
  \label{framework-of-SeqMIA}
\end{figure}

\subsection{Attack Method}
Based on the above, we propose a new membership inference attack, namely the Sequential-metric based Membership Inference Attack (\SeqMIA).
To execute \SeqMIA, the adversary needs to acquire the multi-metric sequences from the training process of the target model. 
However, in this work, we consider the black-box scenario where the adversary can only access the final well-trained target model, i.e., the version at its final training epoch. 
To address this issue, similar to Liu et al.~\cite{liu2022membership}, the adversary leverages knowledge distillation on the target model to obtain its distilled model. 
This way, the adversary gains full control of the distillation process and can save the distilled models at different epochs.
The attacker then evaluates different metrics of a given target sample on each intermediate distillation model to obtain its multi-metric sequence, called \textit{distilled multi-metric sequence}.
Finally, the attack model, functioning as a membership classifier, takes the distilled multi-metric sequence as input to infer the sample's membership status.
The overview of \SeqMIA\ is depicted in \autoref{framework-of-SeqMIA}, involving five stages: shadow model training, model distillation, construction of multi-metric sequences, serialization processing, and membership inference. 

\mypara{Shadow Model Training} 
As mentioned earlier, the adversary holds an auxiliary dataset $D^a$, which follows the same distribution as the target model's training dataset. 
The adversary first divides this auxiliary dataset $D^a$ into two disjoint sets: the shadow dataset $D^s$ and the distillation dataset $D^k$. 
The shadow dataset $D^s$ is divided into two disjoint datasets, namely $D^s_{train}$ and $D^s_{test}$. 
$D^s_{train}$, representing the members, is utilized to train the shadow model, which aims to emulate the behavior of the target model, while $D^s_{test}$ represents the non-members.
Given the assumption specified in \autoref{threatmodel}, the adversary can train a shadow model with the same architecture and hyperparameters of the target model.

\mypara{Model Distillation} 
The distillation dataset $D^k$ is used to distill the target and shadow models, simulating their training process. 
For brevity, we refer to the target model and shadow model as the original models. Following the approach in Liu et al.~\cite{liu2022membership}, we query the two original models to obtain their output posteriors as soft labels and only use $L_{soft}$ (Kullback-Leibler divergence loss between the soft labels and the student model's output posteriors) to train the distilled models for $n$ epochs. 
Subsequently, we capture snapshots of the distilled model's parameters at different epochs, resulting in a series of snapshots $s_1, s_2, ..., s_n$, which mimic the original model's training process. 
Recognizing the significance of membership information contained in the original model's output posteriors, we include the original model as an additional supplement in the snapshots series (denoted as $s_{n+1}$).
While we can obtain the shadow model's training process, it does not match the exact distillation process of the target model. Distilled models converge faster with sufficient distillation data. Consequently, to align the membership information depicted in the training processes of both target and shadow models, we proceed by distilling the shadow model further, aiming to emulate similar training processes.

\begin{figure}[t]
  \centering
  \includegraphics[width=\linewidth]{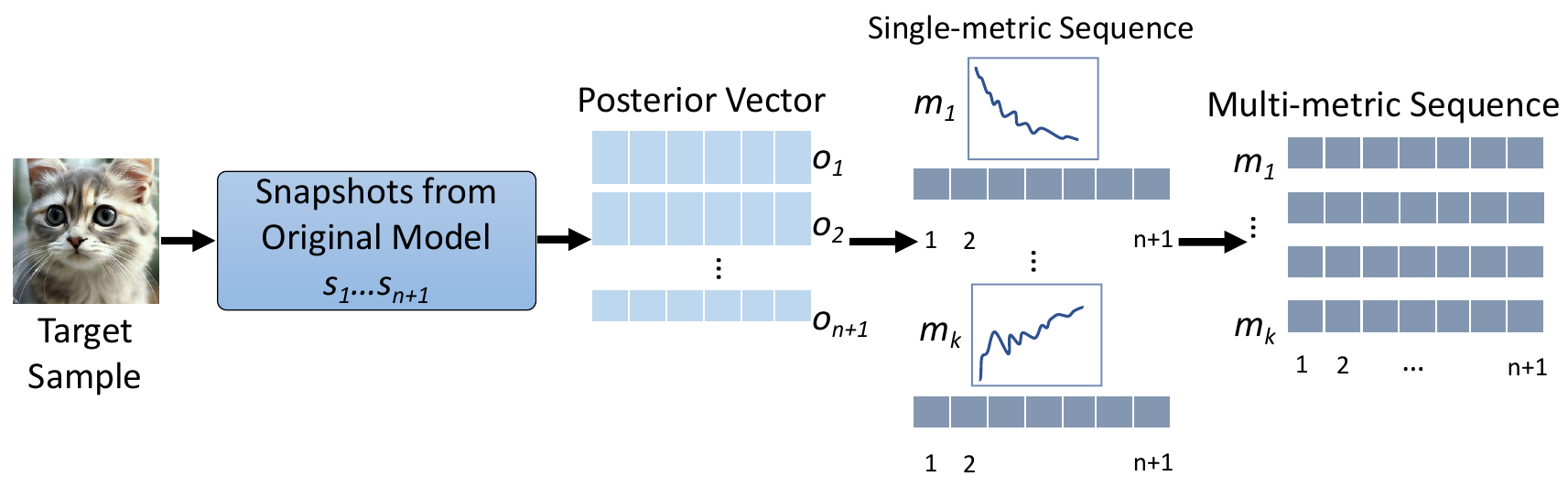}
  \caption{Workflow of multi-metric sequence construction, which assembles the membership information of a sample into a sequence of a $k$-dimensional space.}
  \label{encoding}
\end{figure}

\mypara{Construction of Multi-metric Sequences}
Our construction method involves serialized feature engineering to encode membership information leaked from the output posteriors of $n+1$ snapshots into sequences in a high-dimensional space. 
To achieve this, we feed a given sample into $s_1, s_2, ..., s_n, s_{n+1}$ and obtain $n+1$ output posteriors $o_1, o_2, ..., o_n, o_{n+1}$. 
Subsequently, we calculate $k$ metric values (e.g., Loss, Max, SD, etc., as mentioned in \autoref{metrics}) for each output posterior $o_i$. 
These $n+1$ values of the same metric are concatenated together in temporal order, forming a single metric sequence, which becomes a $(n+1)$-dimensional vector. 
Finally, we concatenate the $k$ metric sequences together as a sequential membership signal in a $k$-dimensional space called \textit{multi-metric sequence}, represented as a $k*(n+1)$ matrix. 
See \autoref{encoding} for an illustration of how to construct the multi-metric sequence.

\mypara{Serialization Processing}
As the shadow model and its distilling process are fully controlled by the adversary, they label the multi-metric sequence obtained from $D^s_{train}$ as 1 (members), and that from $D^s_{test}$ as 0 (non-members). 
Subsequently, the adversary constructs a binary dataset and uses it to train the attack model.
As aforementioned, the MLP model ignores sequential or time-series information in the multi-metric sequence, which may cause the loss of some membership signals. 
Therefore, we utilize a recurrent neural network (RNN) attack model to process this sequential data. 
Specifically, to adequately emphasize the significance of different points in the multi-metric sequence, we employ an attention-based RNN as the attack model. 
This choice allows us to capture contextual semantics by learning weights that highlight key points in such signals for membership inference.
We train the attack model by minimizing the cross-entropy loss for the binary classification task. 

\mypara{Membership Inference} With the trained attack model, the adversary can perform membership inference on a given target sample by following these steps: First, the target sample is encoded into multi-metric sequence by feeding it to the series of snapshots $s_1,s_2,...,s_n,s_{n+1}$ which are from the target model. 
Then, this sequence can be fed into the attack model to predict its membership status, i.e., 1 or 0.

\section{Experimental Setup}

\subsection{Datasets}
We consider seven benchmark datasets of different tasks, sizes, and complexity to conduct our experiments. 
Concretely, we adopt four computer vision datasets, namely CIFAR10~\cite{krizhevsky2009learning}, CIFAR100~\cite{krizhevsky2009learning}, CINIC10~\cite{darlow2018cinic}, GTSRB~\cite{stallkamp2012man}, and three non-computer vision datasets, namely Purchase~\cite{Purchases}, News~\cite{News} and Location~\cite{Locations}.
See details in \autoref{app:dataset}.

\begin{table}
\footnotesize
  \caption{Performance of target models, wherein training/testing accuracy is reported for each model.}
  \label{performance-of-target-models}
  \centering
  \scalebox{0.95}
    {\begin{tabular}{c c c c c }
    \toprule
    Target model &CIFAR10 &CIFAR100 &CINIC10 &GTSRB \\
    \midrule
    VGG-16& 1.000/0.756& 1.000/0.296& 1.000/0.569& 1.000/0.923 \\
    ResNet-56& 0.987/0.662& 0.998/0.243& 0.972/0.472& 1.000/0.930 \\
    WideResNet-32& 0.991/0.710& 0.976/0.371& 0.952/0.502& 0.999/0.912\\
    MobileNetV2& 0.986/0.667& 0.998/0.218& 0.972/0.463& 1.000/0.917\\
    \midrule
    \midrule
    Target model &News &Purchase &Location & \\
    \midrule
    MLPs& 0.976/0.663& 1.000/0.716& 1.000/0.568  \\
    \bottomrule
  \end{tabular}}
\end{table}

Following~\cite{liu2022membership}, we divide each dataset into five parts: target training/testing dataset ($D^t_{train}$ / $D^t_{test}$), shadow training/testing dataset ($D^s_{train}$ / $D^s_{test}$), and distillation dataset $D^k$. Among them, $D^s_{train}$, $D^s_{test}$ and $D^k$ are disjoint subsets of the auxiliary dataset $D^a$ held by the adversary. Specifically, the data partitioning is such that the sizes of the former four datasets are kept exactly the same, and the remaining data samples are placed into the distillation dataset (see details of data splitting in Appendix \autoref{data-splits}). 

\subsection{Models}
For image datasets, we adopt WideResNet-32~\cite{zagoruyko2016wide}, VGG-16~\cite{simonyan2014very}, MobileNetV2~\cite{sandler2018mobilenetv2} and ResNet-56~\cite{he2016deep}, as our target models. 
For the non-image datasets, we adopt a 2-layer MLP as the target model. 
These models are trained from 80 to 150 epochs due to the complexity of model architectures and datasets. For distillation, the epoch number is set to 50. The optimization algorithm used is SGD, with a learning rate ranging from 0.01 to 0.1. 
See the target models' performance in \autoref{performance-of-target-models}.
Lastly, both the architecture of the shadow model and the distilled models in our experiments remain consistent with that of the target model. 
Note that since recent research~\cite{choquette2021label} has shown that data augmentation increases membership leakage, all attack methods in this paper, including ours, are performed on target models without data augmentation.

\begin{figure*}[t]
  \centering
  \begin{subfigure}[t]{\size}
    \centering
    \includegraphics[width=\size]{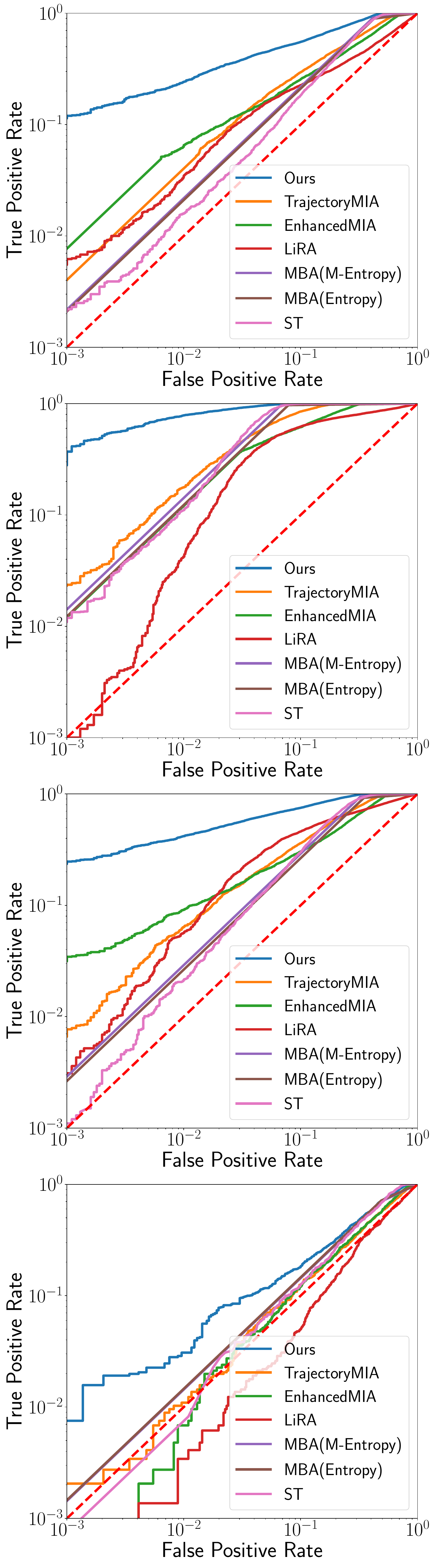}
    \caption{VGG-16}
    \label{ROC-VGG-16}
  \end{subfigure}
  \begin{subfigure}[t]{\size}
    \centering
    \includegraphics[width=\size]{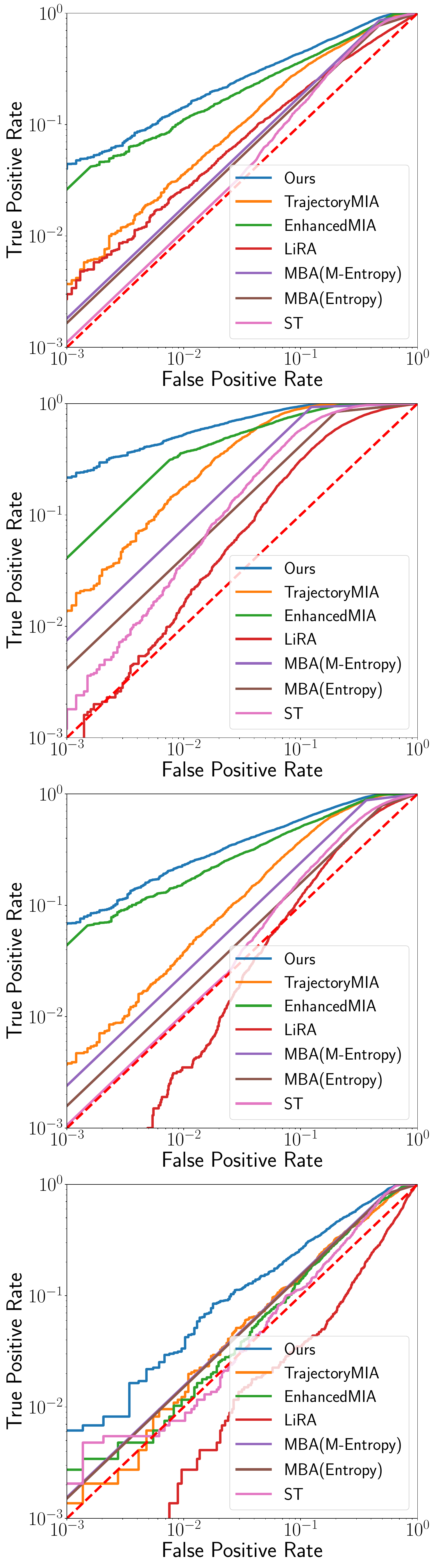}
    \caption{ResNet-56}
    \label{ROC-ResNet-56}
  \end{subfigure}
  \begin{subfigure}[t]{\size}
    \centering
    \includegraphics[width=\size]{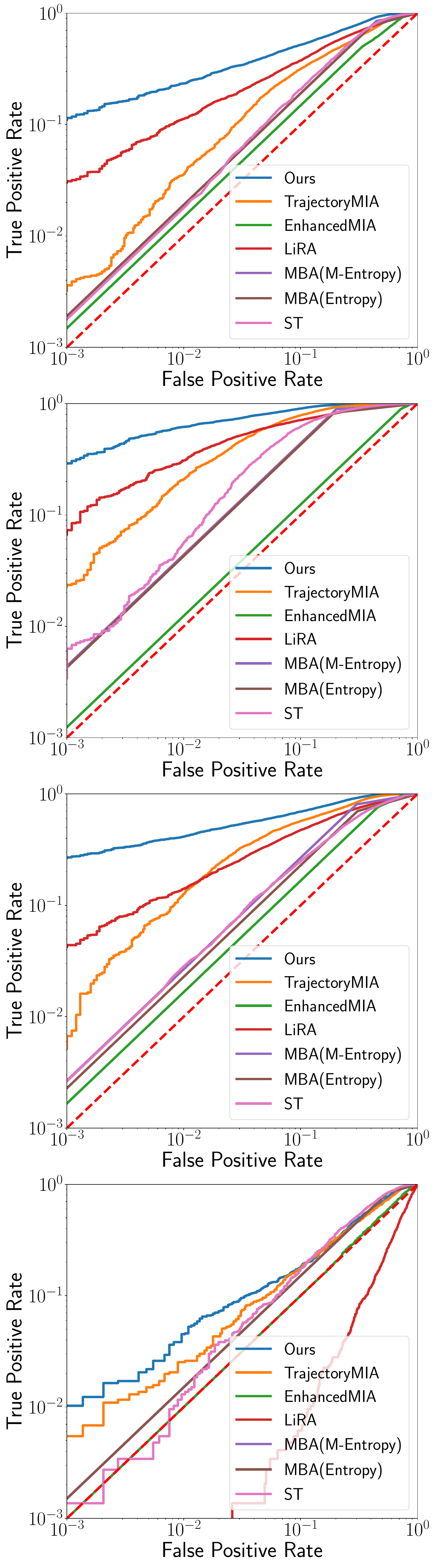}
    \caption{WideResNet-32}
    \label{ROC-WideResNet-32}
  \end{subfigure}
  \begin{subfigure}[t]{\size}
    \centering
    \includegraphics[width=\size]{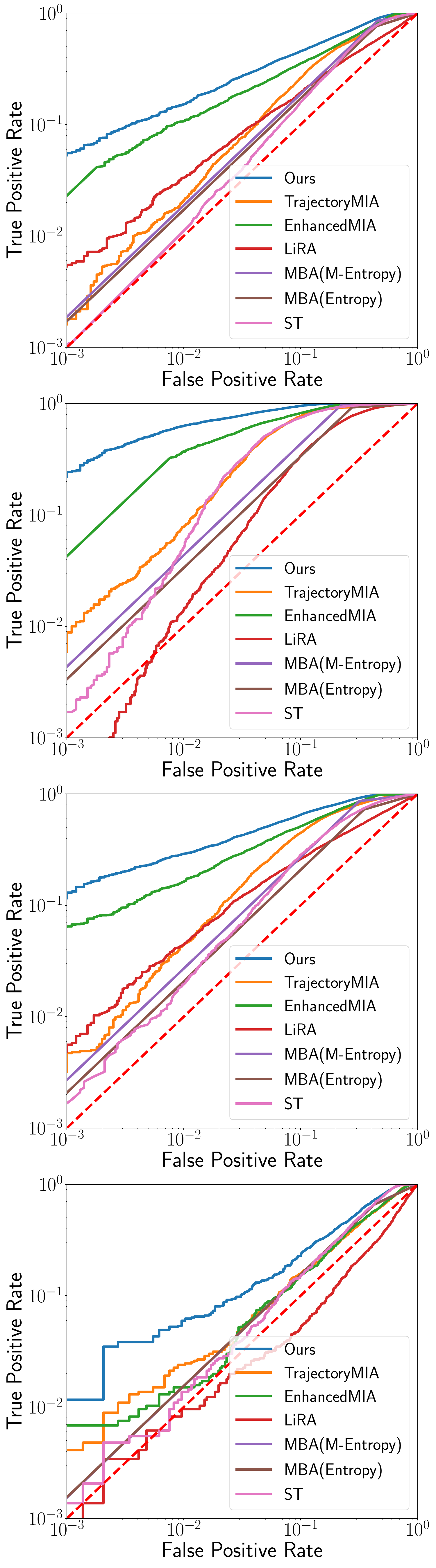}
    \caption{MobileNetV2}
    \label{ROC-MobileNetV2}
  \end{subfigure}
  
  \caption{Log-scale ROC curves for attacks on different model architectures and four image datasets (from top to bottom: CIFAR10, CIFAR100, CINIC10, and GTSRB).}
  \label{ROC-CIFAR100}
\end{figure*}

\subsection{Baselines}
To demonstrate the effectiveness of \SeqMIA, we compared it with the following MIA methods. 

\mypara{Shadow Training} Shadow training is a method proposed by Shokri et al.~\cite{shokri2017membership}, which uses multiple shadow models to mimic the target model and assigns membership labels to the output posteriors from shadow models. With a large number of labeled output posteriors, it is feasible to train an attack model. Further, Salem et al.~\cite{salem2018ml} employ only one shadow model to improve this method and achieve similar attack performance. In this study, the improved shadow training method~\cite{salem2018ml} is adopted as one of our baselines, denoted as ST.

\mypara{Metric-based Attack} Metric-based attack~\cite{song2021systematic} is performed directly based on some metric values calculated from the output posteriors of the target model, and it does not require training the attack model. In this paper, we choose two metrics, prediction entropy, and modified prediction entropy, for the baseline attacks, which are denoted as MBA(Entropy) and MBA(M-Entropy), respectively.

\mypara{LiRA} LiRA~\cite{carlini2022membership} trains $N$ reference models, of which $N/2$ are IN models (trained with the target sample), and $N/2$ are OUT models (trained without the target sample). Then, it calculates the Gaussian distributions of losses on the target sample for IN models and OUT models. Finally, it measures the likelihood of the target sample's loss (output by the target model) under each of the distributions, and returns whichever is more likely (i.e., member or non-member). 
Since the online attack of LiRA requires training new IN models for every (batch of) target samples, we use its offline version in our main evaluation, denoted as LiRA. We also provide a comparison between its online version, denoted as LiRA (online), and our method on the part of datasets and models. 

\mypara{EnhancedMIA} EnhancedMIA~\cite{ye2022enhanced} utilizes $N$ distilled models to capture the loss distribution of the target sample. Since these $N$ distilled models are trained on an auxiliary dataset relabeled with the target model, this approach eliminates the uncertainty with regard to the training set and the target sample.

\mypara{TrajectoryMIA} TrajectoryMIA~\cite{liu2022membership} is another state-of-the-art attack method, which exploits membership information leaked from the training process of the target model. 

\begin{figure*}[h]
  \centering
  \includegraphics[width=0.9\linewidth]{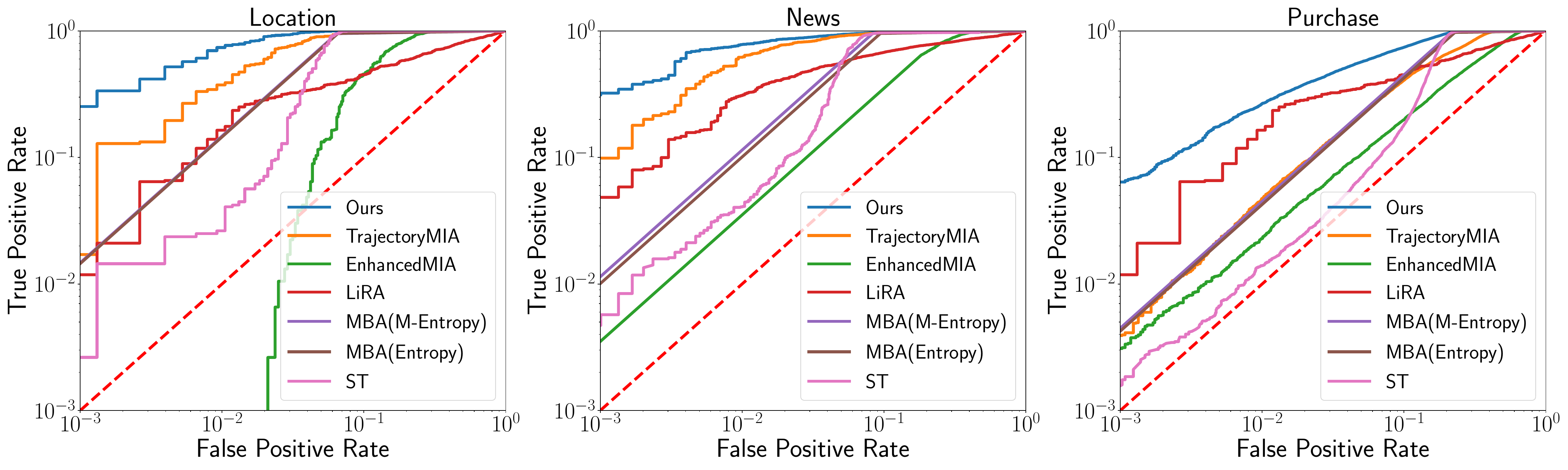}
  \caption{Log-scale ROC curves for attacks against MLPs trained on three non-image datasets.}
  \label{ROC-Non-image-dataset}
\end{figure*}

\begin{table*}[h]
\footnotesize
\caption{Attack performance of different attacks against VGG-16 trained on four image datasets.}
\label{attack-performance-on-VGG-16}
    \centering
    \scalebox{0.85}
    {
    \begin{tabular}{c c c c c c c c c c c c c}
        \toprule
         \multirow{2}{*}{MIA method} & \multicolumn{4}{c}{TPR @ 0.1\% FPR (\%)} &
         \multicolumn{4}{c}{Balanced accuracy} &
         \multicolumn{4}{c}{AUC}\\
         \cmidrule(r){2-5} \cmidrule(r){6-9} 
         \cmidrule(r){10-13} 
          & CIFAR10 & CIFAR100 & CINIC10 & GTSRB & CIFAR10 & CIFAR100 & CINIC10 & GTSRB & CIFAR10 & CIFAR100 & CINIC10 & GTSRB\\
         \midrule
         ST & 0.22 & 1.19 & 0.11 & 0.08 & 0.726 & 0.923 & 0.723 & 0.570 & 0.753 & 0.966 & 0.833 & 0.635\\
         MBA(Entropy) & 0.21 & 1.23 & 0.27 & 0.15 & 0.735 & 0.945 & 0.789 & \textbf{0.613} & 0.735 & 0.945 & 0.788 & 0.613\\
         MBA(M-Entropy) & 0.22 & 1.43 & 0.29 & 0.14 & 0.747 & 0.952 & 0.814 & 0.606 & 0.747 & 0.952 & 0.814 & 0.606\\
         LiRA & 0.62 & 0.10 & 0.31 & 0.00 & 0.565 & 0.775 & 0.707 & 0.508 & 0.575 & 0.819 & 0.756 & 0.480\\
         EnhancedMIA & 0.77 & 1.21 & 3.45 & 0.00 & 0.636 & 0.836 & 0.717 & 0.567 & 0.694 & 0.904 & 0.767 & 0.575\\
         TrajectoryMIA & 0.40 & 2.36 & 0.77 & 0.21 & 0.666 & 0.892 & 0.730 & 0.540 & 0.739 & 0.949 & 0.811 & 0.560\\
         \midrule
         SeqMIA & \textbf{11.99} & \textbf{37.03} & \textbf{24.70} & \textbf{0.75} & \textbf{0.766} & \textbf{0.959} & \textbf{0.850} & 0.577 & \textbf{0.869} & \textbf{0.992} & \textbf{0.937} & \textbf{0.649}\\
         \bottomrule
    \end{tabular}
    }
\end{table*}

Among the aforementioned methods, the first two attacks represent conventional approaches that utilize the output posteriors. LiRA and EnhancedMIA are two SOTA attacks that employ multiple reference/distilled models to calibrate membership information derived from the output posteriors. Additionally, TrajectoryMIA is another SOTA attack that leverages supplementary membership signals in conjunction with the output posterior.

Lastly, when performing LiRA and EnhancedMIA, we follow previous work~\cite{carlini2022membership} and set $N = 64$ for image datasets and $N = 256$ otherwise.

\subsection{Evaluation Metrics} 
First, we adopt two average-case metrics, balanced accuracy and AUC, which have been widely used in~\cite{li2021membership, choquette2021label, nasr2019comprehensive, hayes2017logan, chen2020gan, hilprecht2019reconstruction, hisamoto2020membership}.

\mypara{Balanced accuracy} Balanced accuracy is the probability that a membership inference attack makes a correct prediction on a balanced dataset of members and non-members.

\mypara{AUC} AUC is the area under the receiver operating characteristic (ROC) curve, which is formed by the true-positive rate (TPR) and false-positive rate (FPR) of a membership inference attack for all possible thresholds.

Further, we use TPR @ low FPR and Full log-scale ROC as another two metrics recently proposed by Carlini et al.~\cite{carlini2022membership}. 
This is due to that a reliable inference attack targeting a small number of samples in the entire dataset should be taken seriously. Meanwhile, the TPR at high FPR is unreliable to the adversary. Therefore, these metrics have been used in recent works~\cite{ye2022enhanced, liu2022membership} to evaluate the utility of MIAs more comprehensively.

\mypara{TPR @ low FPR} TPR @ low FPR reports the true-positive rate at a single low false-positive rate (e.g., 0.1\% FPR), which allows for a quick review of the attack performance on a small portion of samples in the entire dataset.

\mypara{Full log-scale ROC} Full log-scale ROC highlights TPRs in low FPR regions by drawing the ROC curves in logarithmic scale, which provides a more complete view of attack performance than TPR @ low FPR.

\section{Experimental Results}

\subsection{Attack Performance}\label{performance}
The attack performance of our \SeqMIA\ and baseline attacks is presented in \autoref{ROC-CIFAR100} and \autoref{ROC-Non-image-dataset}.
First, we observe that \SeqMIA\ achieves the best performance in almost all cases.
Specifically, for TPR @ 0.1\% FPR shown in \autoref{attack-performance-on-VGG-16}, \SeqMIA\ demonstrates an order of magnitude improvement compared to the baseline attacks.
Regarding the two averaged metrics, balanced accuracy, and AUC, we can also find that \SeqMIA\ outperforms all baseline attacks in most cases. Additional results can be found in Appendix \autoref{attack-performance-on-ResNet-56}, \autoref{attack-performance-on-WideResNet-32}, \autoref{attack-performance-on-MobileNetV2} and \autoref{attack-performance-on-MLPs}.

\begin{figure*}[!h]
  \centering
  \includegraphics[width=0.95\linewidth]{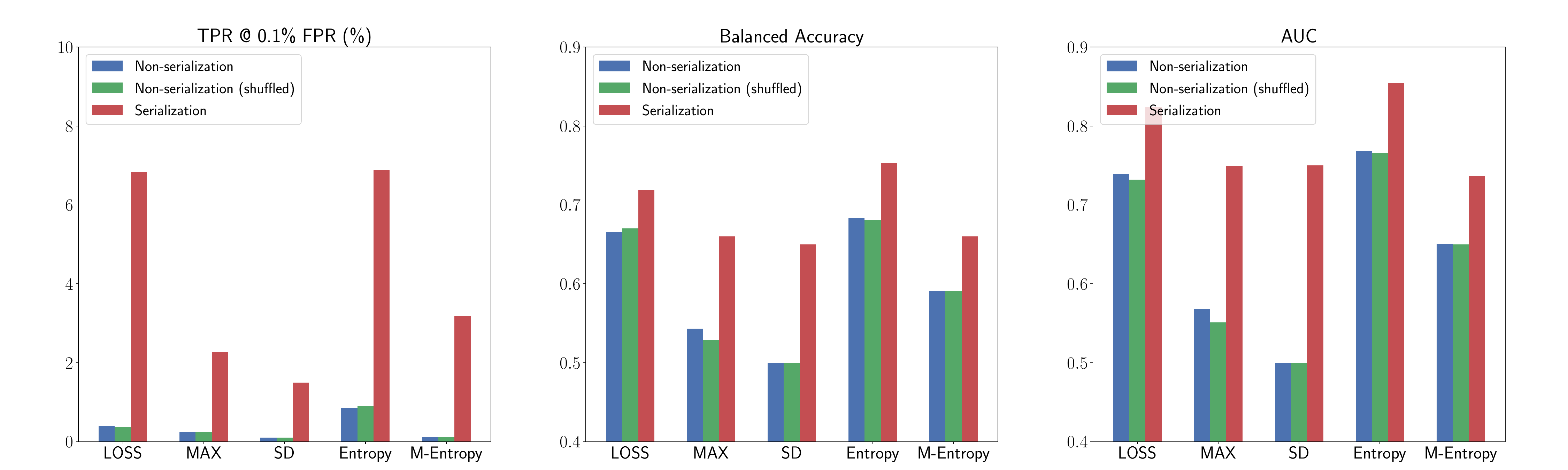}
  \caption{Performance of attacks using serialized and non-serialized membership signals against VGG-16 trained on CIFAR10.}
  \label{single-metric-VGG-16}
\end{figure*}

\begin{table}[h]
\footnotesize
\caption{Attack performance of \SeqMIA\ and LiRA (online) against VGG-16 trained on CIFAR10 and CIFAR100.}
\label{comparion to LiRA(online)}
    \centering
    \scalebox{0.71}
    {
    \begin{tabular}{c c c c c c c c}
        \toprule
         \multirow{2}{*}{MIA method} & \multicolumn{2}{c}{TPR @ 0.1\% FPR (\%)} &
         \multicolumn{2}{c}{Balanced accuracy} &
         \multicolumn{2}{c}{AUC}\\
         \cmidrule(r){2-3} \cmidrule(r){4-5} 
         \cmidrule(r){6-7} 
          & CIFAR10 & CIFAR100 & CIFAR10 & CIFAR100  & CIFAR10 & CIFAR100 \\
          \midrule
          LiRA(online) & 6.38 & 19.91 & 0.687 & 0.884 & 0.766 & 0.961 \\
          SeqMIA & 11.99 & 37.03 & 0.766 & 0.959 & 0.869 & 0.992 \\
         \bottomrule
    \end{tabular}
    }
\end{table}

Furthermore, even on the well-generalized model, \SeqMIA\ exhibits a notable advantage over other baseline attacks in terms of TPR @ 0.1\% FPR.
For instance, VGG-16 trained on GTSRB achieves training and testing accuracies of 1.000 and 0.923, respectively, indicating a well-generalized target model (see \autoref{performance-of-target-models}).
For this model, we surprisingly find that two state-of-the-art attacks, LiRA and EnhancedMIA, only achieve 0\% TPR @ 0.1\% FPR, as shown in \autoref{attack-performance-on-VGG-16}. The state-of-the-art attack, TrajectoryMIA, only achieves a TPR of 0.21\% at an FPR of 0.1\%. In contrast, our \SeqMIA\ demonstrates an impressive 0.75\%. 
This superior performance can be attributed to its ability to capture and leverage the integrated membership signals: \textit{Pattern of Metric Sequence}, even in scenarios where the model is well-generalized.

Lastly, as shown in \autoref{comparion to LiRA(online)}, even when compared to the costly but effective method LiRA(online), \SeqMIA\ consistently outperforms it, especially regarding TPR @ 0.1\% FPR and balanced accuracy. This comparison further emphasizes the effectiveness of \SeqMIA.

\subsection{Analysis}\label{Analysis}
Recall that we use a simple yet effective approach, the attention-based RNN, as our attack model to process the multi-metric sequences. 
This method automatically uncovers various patterns in the metric sequences without requiring prior characterization. 
While our \SeqMIA\ has shown superior performance, we further investigate its success, particularly focusing on whether the attention-based RNN can indeed uncover the different patterns in the metric sequence as claimed.

\mypara{Serialization vs. Non-serialization}
We first investigate whether \SeqMIA\ can indeed distinguish the pattern of metric sequences between members and non-members. To mitigate the effects of multiple metrics, we use only one metric at a time. 
In particular, we consider our \SeqMIA, which utilizes an attention-based RNN attack model to serialize the input, referred to as Serialization. 
For comparison, we also consider TrajectoryMIA, which uses an MLP attack model, referred to as Non-serialization. 
Additionally, we introduce a variant of TrajectoryMIA where the order of the metrics is randomized, denoted as Non-serialization (shuffled).
As shown in \autoref{single-metric-VGG-16}, we can observe that both Non-serialization and Non-serialization (shuffled) achieve similar performance in all cases. These results indicate that the MLP attack model treats the input values as independent and ignores any sequential or time-series information present in the input vector. In contrast, Serialization achieved significantly better attack performance in all cases. These results suggest that attention-based RNNs processing either sequence data or time-series data do discover patterns of metric sequences between members and non-members.

\begin{figure}[t]
  \centering
  \begin{subfigure}[t]{\size}
    \centering
    \includegraphics[width=\size]{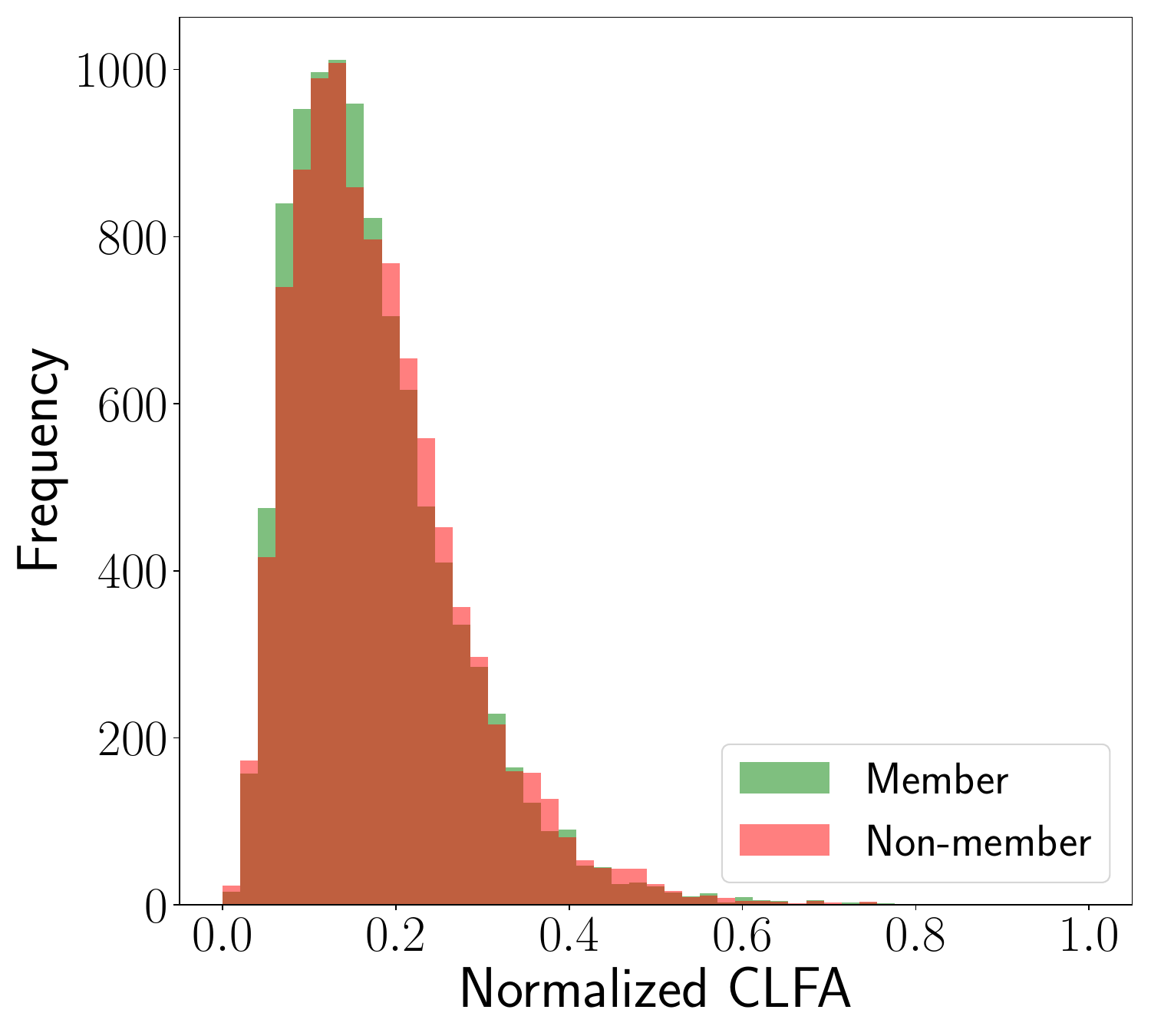}
    \caption{First 10 Epochs}
    \label{FRONT}
  \end{subfigure}
  \begin{subfigure}[t]{\size}
    \centering
    \includegraphics[width=\size]{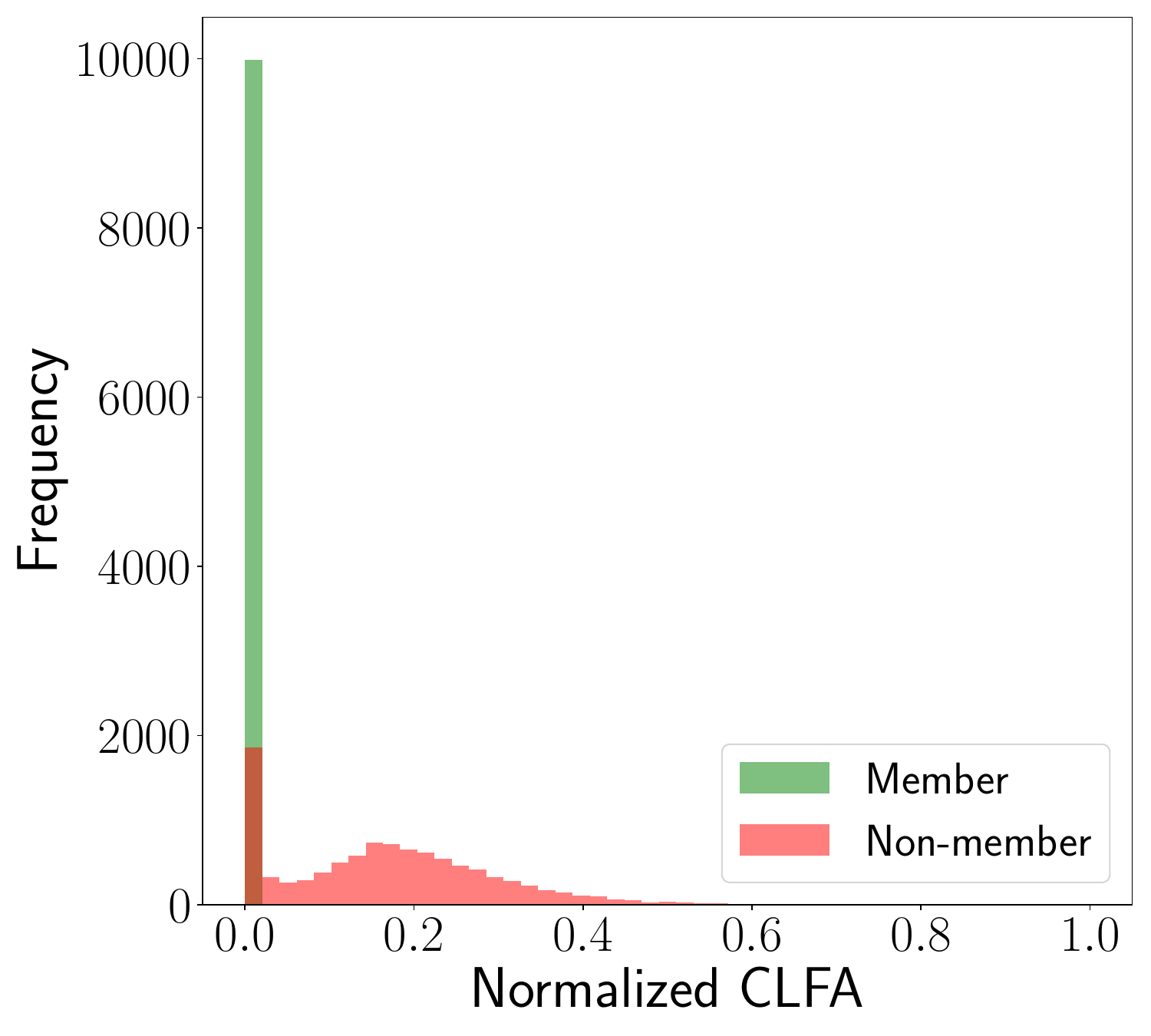}
    \caption{Last 10 Epochs}
    \label{END}
  \end{subfigure}
  \caption{Distribution of CLFA for members and non-members in different training epochs for VGG-16 trained on CIFAR100.}
  \label{fluctuation-amplitude}
\end{figure}

\begin{figure}[t]
  \centering
  \includegraphics[width=0.56\linewidth]{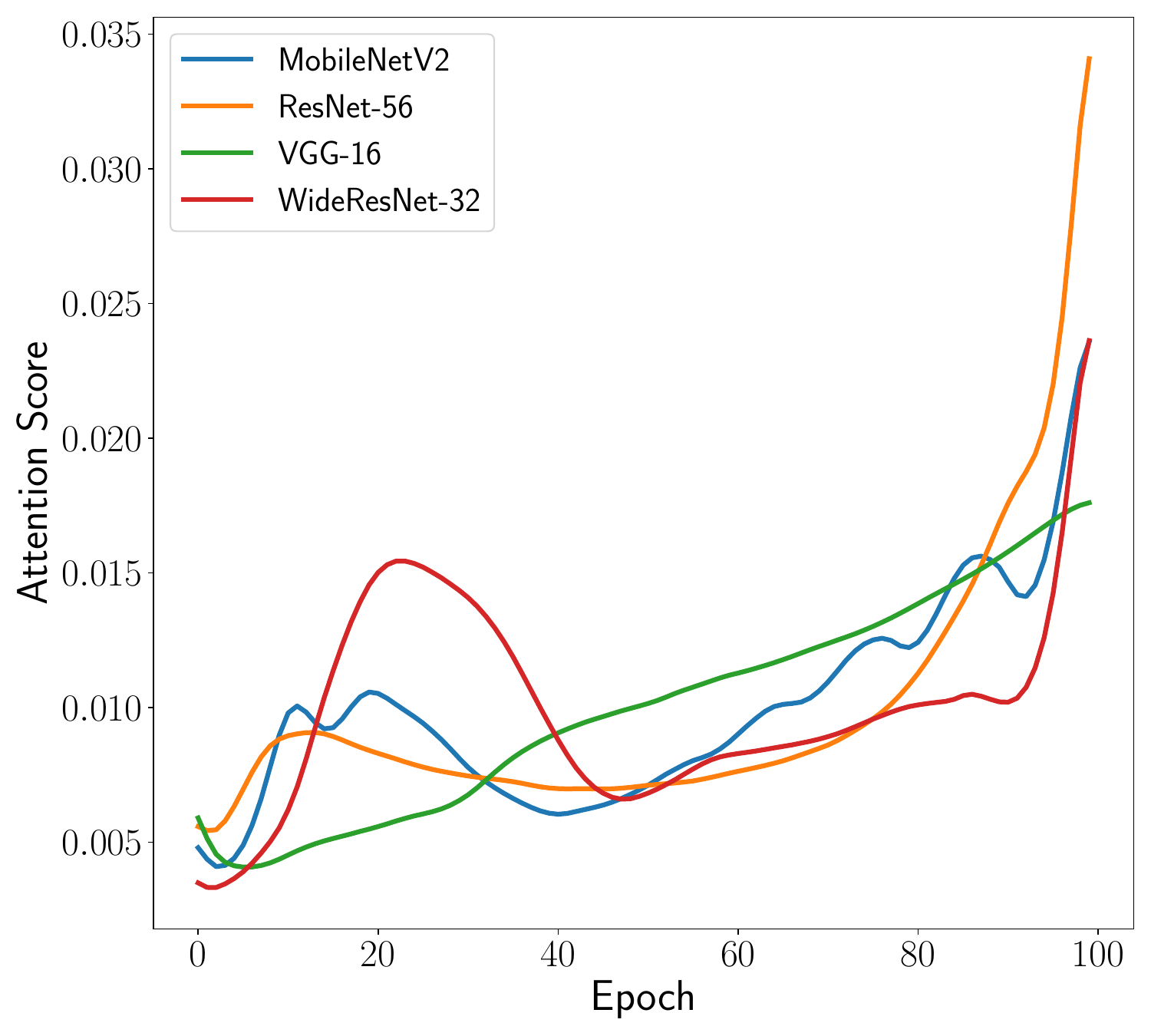}
  \caption{Attention score of our attack against four models trained on CIFAR100.}
  \label{attention_score}
\end{figure}

Now, we further discuss why we should adopt the attention mechanism.
As shown in \autoref{fluctuation-amplitude}, the magnitude of the loss value fluctuations for both members and non-members is large in the first 10 training epochs (i.e., the model is in an underfitting state at this stage), and thus it is difficult to distinguish between them. However, when the model is overfitted or close to being overfitted (the last 10 training epochs), members reduce the magnitude of loss fluctuations, while non-members do not. For example, almost all members have loss fluctuations of less than 0.01, whereas more than half of the non-members have fluctuations of more than 0.01. We believe that this is because, at this stage, the model matches the individual characteristics of the members so that they exhibit consistently small losses. Therefore, we introduce an attention mechanism to focus on key parts of the metric sequence. \autoref{attention_score} describes the attention scores of \SeqMIA\ with the four models trained on CIFAR100, which implies that \SeqMIA\ is able to capture the membership signal in the metric sequences accurately.

\begin{table}[t]
\footnotesize
    \caption{TPR @ 0.1\% FPR of RNN-based SeqMIA, Transformer-based SeqMIA, and TrajectoryMIA against ResNet-56 trained on four image datasets.}
\label{tpr-rnn-vs-transformer}
    \centering
    \scalebox{0.95}
    {
    \begin{tabular}{c c c c}
        \toprule
         \multirow{2}{*}{Dataset} & \multicolumn{3}{c}{TPR @ 0.1\% FPR (\%)}\\
         \cmidrule(r){2-4}
         & RNN & Transformer & TrajectoryMIA\\
         \midrule
         CIFAR10 & \textbf{4.43} & 2.55 & 0.37 \\
         CIFAR100 & \textbf{21.67} & 0.65 & 1.38 \\
         CINIC10 & \textbf{6.89} & 5.94 & 0.38 \\
         GTSRB & \textbf{0.62} & 0.62 & 0.14 \\
         \bottomrule
    \end{tabular}
    }
\end{table}

\begin{table}[!t]
\footnotesize
\caption{TPR @ 0.1\% FPR for attacks using different membership signals against VGG-16 trained on four image datasets.}
\label{ablation-study-on-VGG-16}
    \centering
    \scalebox{0.91}
    {
    \begin{tabular}{c c c c c}
        \toprule
         \multirow{2}{*}{MIA method} & \multicolumn{4}{c}{TPR @ 0.1\% FPR (\%)}\\
         \cmidrule(r){2-5} 
          & CIFAR10 & CIFAR100 & CINIC10 & GTSRB\\
         \midrule
         Loss set & 0.40 & 2.36 & 0.77 & 0.21\\
         Multi-metric set & 3.92 & 3.09 & 1.12 & 0.21\\
         \midrule
         Loss sequence & 6.83 & 25.75 & 16.33 & 0.14\\
         Multi-metric sequence & \textbf{11.99} & \textbf{37.03} & \textbf{24.70} & \textbf{0.75}\\
         \bottomrule
    \end{tabular}
    }    
\end{table}

In addition to RNNs, we further explore applying Transformer~\cite{NIPS2017_3f5ee243}, a self-attention-based technique, for serializing metric sequences in \SeqMIA. Transformers allow parallel processing of input sequences, offering efficiency and scalability compared to the sequential processing of RNNs. However, in our evaluation comparing RNN-based and Transformer-based attack models for \SeqMIA, surprisingly, the RNN-based model performs better. We attribute this to sequence length constraints, smaller point dimensions, and the limited amount of sequences, hindering the Transformer's performance. Despite this, the Transformer-based model in \SeqMIA\ generally surpasses TrajectoryMIA, showcasing the effectiveness of serialization in capturing membership information, as shown in \autoref{tpr-rnn-vs-transformer}. See more results in Appendix \autoref{acc-auc-rnn-vs-transformer}.

\mypara{More Signals of Multiple Metric Sequences}
The previous studies~\cite{nasr2019comprehensive,del2022leveraging} have demonstrated the potential of utilizing multiple metrics to enhance performance. However, these studies focus on non-serialized metric values in white-box scenarios and do not consider the influence of serializing multiple metrics. 
Here, we delve deeper into the impact of extra information from multi-metric sequence, which is initially proposed by \SeqMIA.

First, we take TrajectoryMIA as the example (denoted as \textit{Loss Set}) and extend it with multiple metrics (including loss and other metrics in \autoref{metrics}), denoted as \textit{Multi-metric Set}. Both approaches use a set of non-serialized metric values, which is constructed into a vector and fed into an MLP attack model for inference. 
Besides, we denote our \SeqMIA\ as \textit{Multi-metric Sequence}, and its single metric version as \textit{Loss Sequence}. 
As shown in \autoref{ablation-study-on-VGG-16}, the multi-metric set demonstrates higher attack performance than the loss set, which is consistent with the conclusions in ~\cite{nasr2019comprehensive,del2022leveraging}. 
Interestingly, the multi-metric sequence exhibits a larger performance gain compared to the improvement achieved by the multi-metric set. 
For instance, when evaluating VGG-16 trained on CIFAR100, the multi-metric sequence achieves a notable 34.67\% TPR @ 0.1\% FPR improvement over the loss set, while the multi-metric set only improves by 0.73\%. 
See more results in Appendix \autoref{BA-and-AUC-ablation-study-on-VGG-16} and \autoref{BA-and-AUC-ablation-study-on-MLPs}. 
We attribute this to the fact that target models are optimized for member samples such that members will get better on multiple metrics simultaneously, whereas non-members do not. 

\begin{figure*}[t]
  \centering
  \includegraphics[width=0.85\linewidth]{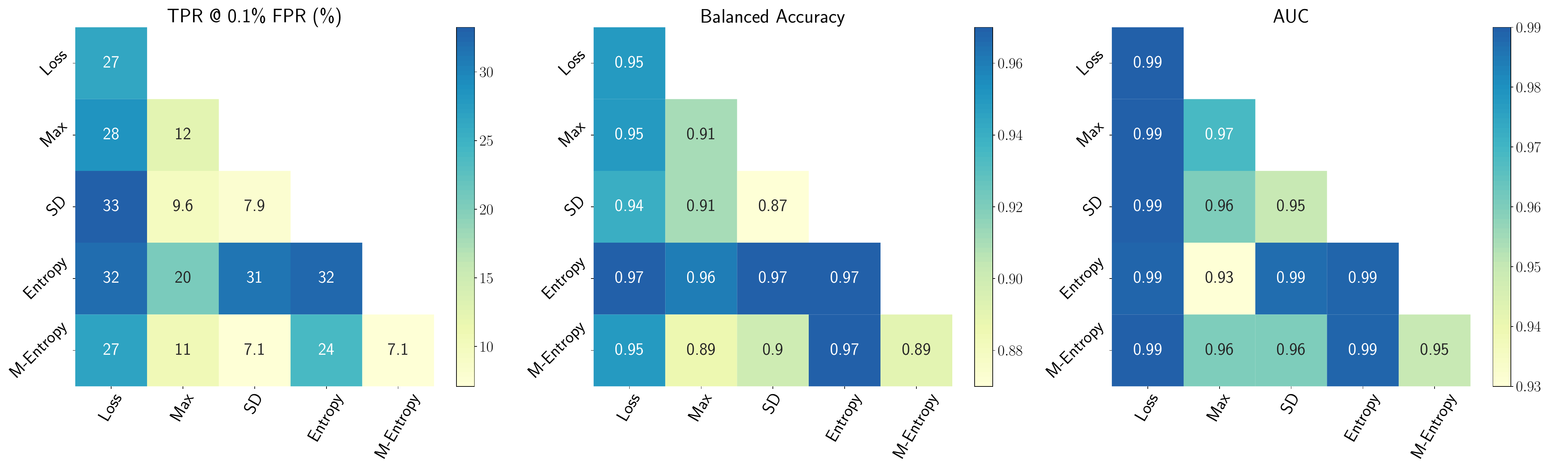}
  \caption{Attack performance of SeqMIA using double metrics against MLPs trained on Location.}
  \label{MLPs-Location-double-metrics}
\end{figure*}

To validate this hypothesis, we calculate the correlation matrix of multi-metric sequences (as depicted in \autoref{metric-correlations}). We can observe that the correlation coefficients for members are usually greater than for non-members. Furthermore, we evaluate the attack performance of \SeqMIA\ using dual-metric sequence, as shown in \autoref{MLPs-Location-double-metrics}. We can find that the best attack performance is often achieved by the two metrics, which have large differences in correlation coefficients between members and non-members. 
For instance, the correlation coefficient of Loss and SD for members is 0.81, whereas, for non-members, it is 0.28, as demonstrated in \autoref{metric-correlations}. 
Meanwhile, Loss and SD achieves the best performance(33\% TPR @ 0.1\% FPR), as shown in \autoref{MLPs-Location-double-metrics}. See more results in Appendix \autoref{VGG-16-CIFAR10-double-metrics}. 
Therefore, we argue that multiple metric sequences indeed contain additional membership information than single metric sequences and can further improve the attack performance.

\subsection{Ablation Study}\label{ablationstudy}
In this section, we investigate the impact of several important factors on the attack performance of our method.

\begin{table}[!t]
\footnotesize
\caption{The impact of the number of distillation epochs for VGG-16 trained on CIFAR10.}
\label{impact-of-number-of-epochs}
    \centering
    \scalebox{0.85}
    {
    \begin{tabular}{c c c c c c c}
        \toprule
         \multirow{2}{*}{} & \multicolumn{6}{c}{Distillation epochs}\\
         \cmidrule(r){2-7} 
          & 5 & 10 & 20 & 30 & 40 & 50\\
         \midrule
         TPR @ 0.1\% FPR (\%)& 1.99 & 5.75 & 9.83 & 10.60 & 12.37 & 11.99\\
         Balanced accuracy & 0.752 & 0.762 & 0.765 & 0.760 & 0.766 & 0.766\\
         AUC & 0.829 & 0.857 & 0.864 & 0.865 & 0.868 & 0.869\\
         \bottomrule
    \end{tabular}
    }
\end{table}

\mypara{Number of Distillation Epochs}
The number of epochs utilized for knowledge distillation significantly impacts both the computational cost in the distillation process and the input dimension to the attack model. 
Therefore, it is crucial to determine the optimal number of epochs required in the distillation process.

\autoref{impact-of-number-of-epochs} illustrates the impact of the number of distillation epochs on the attack performance. 
It is evident that increasing the number of distillation epochs can significantly increase the TPR @ 0.1\% FPR, while having minimal effect on the balanced accuracy and AUC. 
This observation suggests that our attack is capable of distinguishing between members and non-members more reliably. As argued in~\cite{carlini2022membership}, average metrics are often uncorrelated with low FP success rates. While the two average metrics (balanced accuracy and AUC) of our attack no longer grow significantly after 20 distillation epochs, the TPR @ 0.1\% FPR continues to improve. The continued improvement of TPR @ 0.1\% FPR suggests that on a small portion of samples in the entire dataset, our attack becomes more reliable. This situation should be taken into account by the model stakeholders.
Additionally, the best attack performance is achieved within approximately 50 epochs, indicating that the computational cost can be effectively controlled within an acceptable range.

\begin{table}[!t]
\footnotesize
    \caption{The impact of distillation dataset size for VGG-16 trained on CIFAR10. The accuracy of target model is 0.569.}
\label{impact-of-distillation-dataset-size}
    \centering
    \scalebox{0.85}
    {
    \begin{tabular}{c c c c c c c}
        \toprule
         \multirow{2}{*}{} & \multicolumn{6}{c}{Distillation dataset size}\\
         \cmidrule(r){2-7} 
          & 10k & 20k & 70k & 120k & 170k & 220k\\
         \midrule
         Distilled accuracy &  0.564 & 0.566 & 0.563 & 0.561 & 0.574 & 0.566\\
         \midrule
         TPR @ 0.1\% FPR (\%)& 9.38 & 14.83 & 18.22 & 20.96 & 20.98 & 22.57\\
         Balanced accuracy & 0.835 & 0.839 & 0.845 & 0.843 & 0.844 & 0.843\\
         AUC & 0.922 & 0.927 & 0.933 & 0.934 & 0.935 & 0.936\\
         \bottomrule
    \end{tabular}
    }
\end{table}

\mypara{Size of Distillation Dataset}
For knowledge distillation, the size of the distillation dataset is a crucial factor that significantly impacts the distillation performance. 
To investigate the influence of this factor on our attack performance, we conduct experiments with varying sizes of the distillation dataset.

We present the results in \autoref{impact-of-distillation-dataset-size}. Similarly, we observe that a larger distillation dataset size leads to higher TPR @ 0.1\% FPR, while having little impact on the balanced accuracy and AUC. 
This finding demonstrates that a larger distillation dataset is advantageous in improving the attack performance.
Besides, it further supports the claim that our attack becomes very reliable on a small portion of samples in the entire dataset as the size of distillation dataset increases.

\begin{table}[!t]
\footnotesize
    \caption{The impact of the overfitting level of the target model. The experiments are conducted on VGG-16 trained on CINIC10. }
\label{impact-of-overfitting-level}
    \centering
    \scalebox{0.94}
    {
    \begin{tabular}{c c c c c c}
        \toprule
         \multirow{2}{*}{} & \multicolumn{5}{c}{Training dataset size}\\
         \cmidrule(r){2-6} 
          & 30k & 25k & 20k & 15k & 10k\\
         \midrule
         Overfitting level &  0.335 & 0.359 & 0.372 & 0.408 & 0.431\\
         \midrule
         TPR @ 0.1\% FPR (\%)& 8.41 & 10.21 & 11.85 & 13.03 & 20.43\\
         Balanced accuracy & 0.744 & 0.766 & 0.776 & 0.793 & 0.855\\
         AUC & 0.844 & 0.867 & 0.879 & 0.893 & 0.943\\
         \bottomrule
    \end{tabular}
    }
\end{table}

\mypara{Overfitting Level of the Target Model}
It is widely acknowledged that the success of membership inference attacks is closely related to the overfitting level of the target model~\cite{shokri2017membership, salem2018ml}. 
Here we quantify the overfitting level using the training and testing accuracy gap and manipulate it by varying the size of the training set.
Concretely, the distillation dataset size is kept fixed at 100,000 samples, while we manipulate the size of the target/shadow training and testing datasets, ranging from 30,000 down to 10,000 samples.

As described in \autoref{impact-of-overfitting-level}, we observe that as the overfitting level increases, the attack performance improves regarding TPR @ 0.1\% FPR, balancing accuracy, and AUC.
Furthermore, we highlight that even when the target model exhibits good generalization with a low overfitting level (0.335), \SeqMIA\ still achieves a significant 8.41\% TPR @ 0.1\% FPR. 
Surprisingly, this performance outperforms that of all baselines, even in the more overfitting scenario (overfitting level of 0.431), as demonstrated in \autoref{attack-performance-on-VGG-16}.

\begin{figure*}[t]
  \centering
  \includegraphics[width=0.85\linewidth]{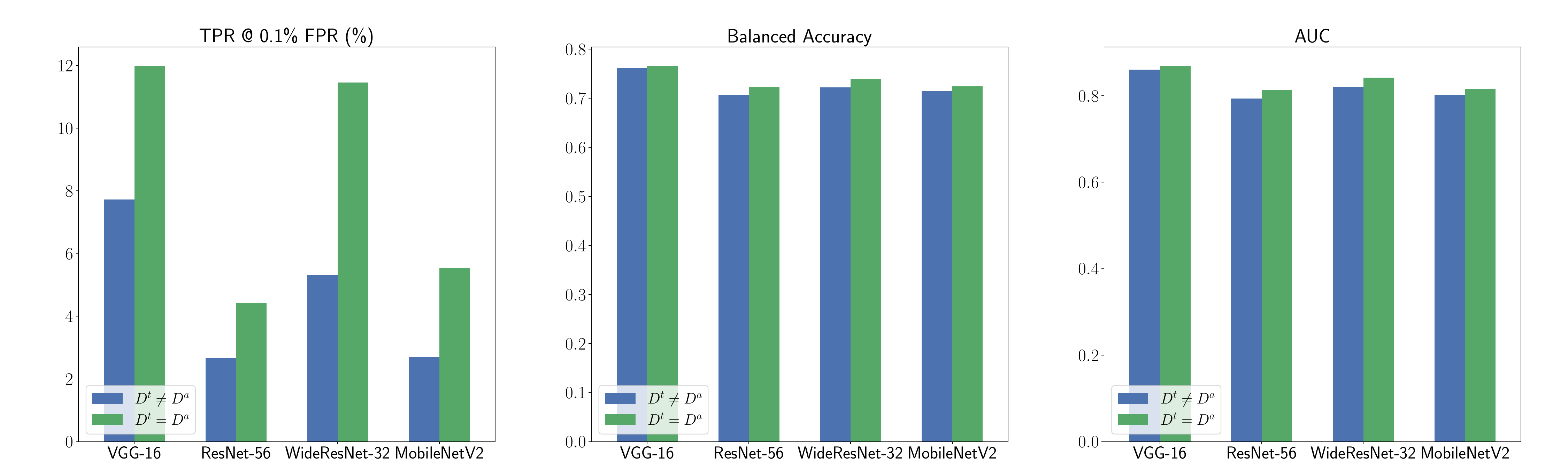}
  \caption{Attack performance of SeqMIA against different models trained on CIFAR10 by using ImageNet part ($D^t \neq D^a$) and CIFAR10 part ($D^t = D^a$) of CINIC10.}
  \label{transfer-attack-with-different-datasets}
\end{figure*}

\begin{figure*}[!t]
  \centering
  \includegraphics[width=0.85\linewidth]{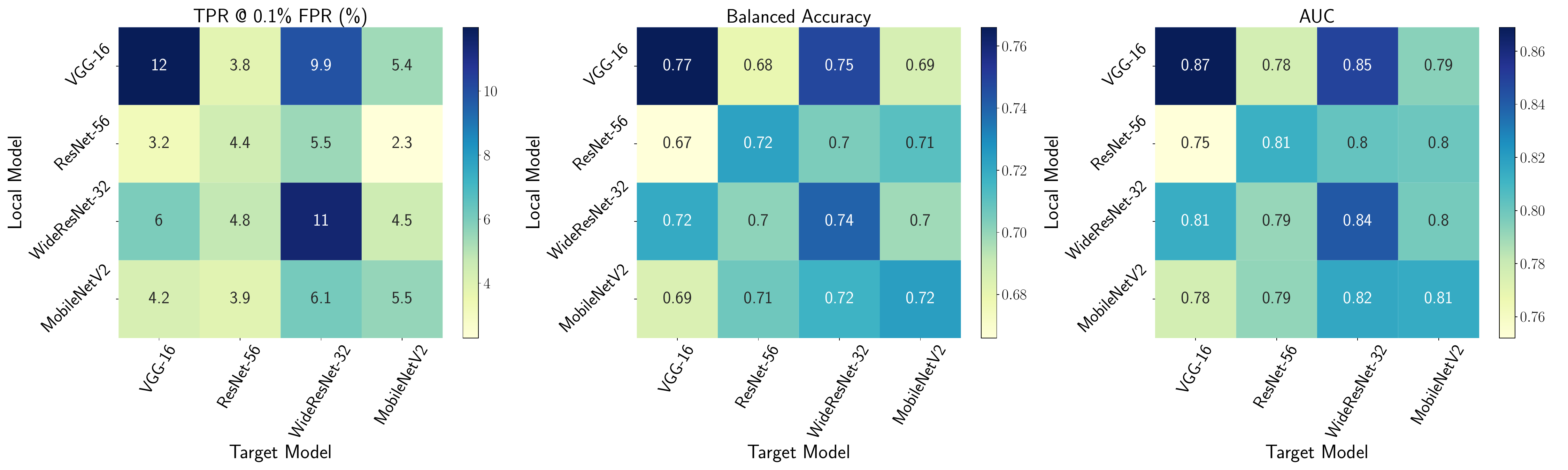}
  \caption{
  The impact of different model architectures used for the local (shadow and distilled) models. The target models and local models are trained on CIFAR10.}
  \label{transfer-attack-with-different-architectures}
\end{figure*}

\mypara{Disjoint Datasets}
We relax our previous assumption that the adversary possesses knowledge of the target model's training dataset distribution. Instead, we utilize CIFAR10 as the training dataset for the target model (denoted as $D^t$) and a subset of CINIC10 derived from ImageNet as the dataset held by the adversary (denoted as $D^a$). 
In \autoref{transfer-attack-with-different-datasets}, we observe that when $D^t \neq D^a$, the attack performance of \SeqMIA\ is compromised. This degradation occurs because the discrepancy in data distribution leads to differences in prediction behavior between the target model and the models trained by the adversary, consequently affecting the quality of our constructed multi-metric sequences in capturing membership information. 
Nonetheless, \SeqMIA's performance still surpasses that of all baselines. 
For instance, \SeqMIA ($D^t \neq D^a$) achieves a TPR @ 0.1\% FPR of more than 7\% against VGG-16 (as shown in \autoref{transfer-attack-with-different-datasets}), while all baselines ($D^t = D^a$) achieve at most 0.77\% (as referred to in \autoref{attack-performance-on-VGG-16}).

\begin{table}[!t]
\footnotesize
\caption{TPR @ 0.1\% FPR of SeqMIA against VGG-16 trained on CIFAR10 with DP-SGD.}
\label{attack-performance-under-dpsgd}
    \centering
    \scalebox{0.9}
    {
    \begin{tabular}{c c c c c}
        \toprule
          & \multicolumn{2}{c}{$\delta=1\textnormal{e-}5$ and $C=1$} &
         \multirow{2}{2cm}{Accuracy of the target model} & \multirow{2}{2cm}{TPR @ 0.1\% FPR (\%)}\\
         \cmidrule(r){2-3}
          & $\sigma$ & $\epsilon$ &  &  \\
         \midrule
         No defense & - & - & 0.756 & 11.99\\
         \midrule
         \multirow{4}{*}{DP-SGD} & 0 & $\infty$ & 0.575 & 0.76\\
          & 0.2 & 1523 & 0.581 & 0.31\\
          & 0.5 & 43 & 0.482 & 0.21\\
          & 1 & 6 & 0.377 & 0.17\\
         \bottomrule
    \end{tabular}
    }
\end{table}

\mypara{Different Model Architectures and Hyperparameters}
We proceed to relax the second assumption that requires the adversary to possess knowledge of the target model's architecture and hyperparameters. 
In other words, the adversary is now allowed to utilize different model architectures and hyperparameters to locally train the shadow model and distilled model.
As depicted in \autoref{transfer-attack-with-different-architectures}, the attack performance is typically optimal along the diagonal. 
This can be attributed to the fact that using the same model architecture and hyperparameters enables the adversary to more accurately simulate the training process of the target model. 
While adopting a different model architecture with different hyperparameters leads to a decrease in \SeqMIA's performance, its worst-case performance still surpasses that of all baselines (when both the target and adversary models share the same architecture and hyperparameters). 
For instance, the worst TPR @ 0.1\% FPR achieved by \SeqMIA\ against VGG-16 is 3.2\% (as shown in \autoref{transfer-attack-with-different-architectures}), while all baselines achieve at most 0.77\% (as referred to in \autoref{attack-performance-on-VGG-16}).

\section{Discussion}

In this section, we evaluate the performance of \SeqMIA\ against several existing defenses. Then, we discuss the limitations of \SeqMIA.

\subsection{Defense Evaluation}\label{Defense}
To mitigate the risk of membership leakage, a large body of defense mechanisms have been proposed in the literature~\cite{li2021membershipMixupMMD, abadi2016deep,zhang2018differentially, pichapati2019adaclip, xu2020adaptive,nasr2018machine,jia2019memguard}. 
In this section, we thoroughly evaluate the effectiveness of \SeqMIA\ against three prominent defenses, namely DP-SGD~\cite{abadi2016deep}, Adversarial Regularization~\cite{nasr2018machine}, and MixupMMD~\cite{li2021membershipMixupMMD}.

\begin{table}[t]
\footnotesize
\caption{TPR @ 0.1\% FPR of different attacks against VGG-16 trained on CIFAR10 with DP-SGD ($\sigma=0.2$ and $\sigma=0.5$).}
\label{attack-baselines-dpsgd}
    \centering
    \scalebox{0.95}
    {
    \begin{tabular}{c c c}
        \toprule
          & \multicolumn{2}{c}{TPR @ 0.1\% FPR (\%)}\\
         \cmidrule(r){2-3} 
          & $\sigma=0.2$ & $\sigma=0.5$ \\
         \midrule
         ST & 0.11 & 0.04 \\
         MBA(Entropy) & 0.11 & 0.10 \\
         MBA(M-Entropy) & 0.11 & 0.10 \\
         LiRA & 0.02 & 0.04\\
         EnhancedMIA & 0.10 & 0.12\\
         TrajectoryMIA & 0.15 & 0.16\\
         \midrule
         SeqMIA & \textbf{0.31} & \textbf{0.21}\\
         \bottomrule
    \end{tabular}
    }
    
\end{table}

\mypara{DP-SGD} 
\autoref{attack-performance-under-dpsgd} presents the performance of \SeqMIA\ under DP-SGD, evaluated on VGG-16 trained on CIFAR10 (see more results in Appendix \autoref{BA-and-AUC-under-dpsgd}). 
We employ the Opacus library to implement DP-SGD and fix the parameters $\delta=1\textnormal{e-}5$ and $C=1$, following~\cite{liu2022membership, carlini2022membership}. 
We can observe that the attack performance of \SeqMIA\ gradually decreases as the defense effects increase. 
However, stronger defense effects also result in a sharp drop in the accuracy of the target model. 
To balance defense strength and model accuracy, we select the cases of $\sigma = 0.2$ and $\sigma = 0.5$ for further analysis. 
These settings offer acceptable trade-offs between defense strength and model accuracy. 
\autoref{attack-baselines-dpsgd} shows that \SeqMIA\ still outperforms other baselines under DP-SGD. 
For instance, when $\sigma = 0.2$, the TPR @ 0.1\% FPR of \SeqMIA\ is more than twice that of other baselines.

\begin{table}[!t]
\footnotesize
    \caption{TPR @ 0.1\% FPR of different attacks against ResNet-56 trained on CINIC10 with AdvReg and MixupMMD.}
\label{attack-against-defense}
    \centering
    \scalebox{1.0}
    {
    \begin{tabular}{c c c c}
        \toprule
         \multirow{2}{*}{} & \multicolumn{3}{c}{TPR @ 0.1\% FPR (\%)}\\
         \cmidrule(r){2-4}
         & no defense & AdvReg & MixupMMD\\
         \midrule
         ST & 0.11 & 1.08 & 0.06\\
         MBA(Entropy) & 0.16 & 0.12 & 0.25\\
         MBA(M-Entropy) & 0.24 & 0.52 & 0.23\\
         LiRA & 0.00 & 0.02 & 0.11\\
         EnhancedMIA & 4.39 & 2.27 & 0.38\\
         TrajectoryMIA & 0.38 & 3.08 & 1.81\\
         \midrule
         SeqMIA & \textbf{6.89} & \textbf{24.62} & \textbf{4.62}\\
         \bottomrule
    \end{tabular}
    }
\end{table}

\mypara{Adversarial Regularization} 
Adversarial Regularization (AdvReg) is an adversarial training-based defense that adds noise to the output posteriors, making it challenging for adversaries to distinguish between members and non-members. 
As demonstrated in \autoref{attack-against-defense} (see more results in Appendix \autoref{BA-and-AUC-against-defense}), \SeqMIA\ continues to achieve the best attack performance in almost all cases.
Interestingly, we observe that AdvReg's co-training with the target model results in members being more involved in the training of the target model, which makes them more significantly different from non-members.
Thus, both \SeqMIA\ and TrajectoryMIA demonstrate enhanced attack performance. Notably, this enhancement is more pronounced for \SeqMIA, as it leverages more membership signals leaked from the training process. 
For instance, when the target model has no defense, \SeqMIA\ achieves a TPR @ 0.1\% FPR of 6.89\%, and when the target model is protected by AdvReg, the TPR @ 0.1\% FPR increases to 24.62\%.

\mypara{MixupMMD} MixupMMD is a defense aimed at mitigating membership inference attacks by reducing the target model's generalization gap. 
As previously discussed, the overfitting level of the target model plays a crucial role in membership leakage. 
Consequently, MixupMMD leads to a degradation in the performance of all attacks, including our \SeqMIA, as depicted in \autoref{attack-against-defense} (see more results in Appendix \autoref{BA-and-AUC-against-defense}). 
However, it is worth noting that despite this degradation, \SeqMIA\ continues to outperform other baseline attacks in almost all cases.

\subsection{Limitations}

\SeqMIA\ has limitations as follows: it cannot be applied to label-only scenarios due to its reliance on the output posterior, and it is not suitable for large model scenarios regarding computation because it requires training and distilling the shadow model. Therefore, model holders can only provide predicted labels instead of the posterior to defend against \SeqMIA.

\section{Related Works}

\subsection{Membership Inference Attacks}
Nowadays, there exist a wide range of other security and privacy research in the machine learning domain~\cite{li2019prove,sha2023fake,wu2022membership,mei2023notable,yang2023data,he2022membership,li2023unganable,liu2023backdoor,shen2022backdoor,zhang2023securitynet,han2022fuzzgan,han2024detection,chen2023comprehensive,liu2023watermarking,ma2023generative,wu2024quantifying,wen2023adversarial,salem2022dynamic}.
In this work, we mainly focus on membership inference attacks.
Membership inference attacks have been successfully performed in various settings about the adversary’s knowledge, including white-box~\cite{nasr2019comprehensive, leino2020stolen}, black-box~\cite{shokri2017membership, salem2018ml, song2021systematic, zhang2021membership, hisamoto2020membership, chen2021machine}, and label-only~\cite{li2021membership, choquette2021label} settings. They have been applied in many machine learning scenarios, such as federated learning~\cite{nasr2019comprehensive, melis2019exploiting, truex2019demystifying} and multi-exit networks~\cite{li2022CCS}, etc. 

Specifically, Shokri et al.~\cite{shokri2017membership} and Salem et al.~\cite{salem2018ml} proposed a shadow training technique that employs shadow models to acquire the membership signals. Moreover, Song et al.~\cite{song2021systematic} and Yeom et al.~\cite{yeom2018privacy} proposed the metric-based attack that directly compares losses or other metric values of samples with a predefined threshold. In addition, some membership signals obtained in the white-box scenario are incorporated to improve the attack performance~\cite{nasr2019comprehensive, del2022leveraging}.
Besides, label-only attacks~\cite{li2021membership, choquette2021label, xu2023membership} solely rely on the predicted labels to acquire the membership signals. Recently, researchers~\cite{sablayrolles2019white, watson2021importance, carlini2022membership, ye2022enhanced, liu2022membership} focused on reducing the false positives of MIAs by using each sample's hardness threshold to calibrate the loss from the target model. Further, Bertran et al.~\cite{bertran2024scalable} proposed a new attack via quantile regression, which can obtain performance close to that of LiRA~\cite{carlini2022membership} with less computation.
Moreover, Liu et al.~\cite{liu2022membership} presented TrajectoryMIA, which utilizes the membership signals generated during the training of the target model.

\subsection{Defenses Against MIAs}
Since the overfitting level is an important factor affecting membership leakage, some regularization techniques have been used by~\cite{shokri2017membership, salem2018ml, liu2022ml} to defend against membership inference attacks, such as L2 regularization, dropout and label smoothing, etc. Recently, Li et al.~\cite{li2021membershipMixupMMD} proposed the method MixupMMD to mitigate membership inference attacks by reducing the target model’s generalization gap. Furthermore, Abadi et al.~\cite{abadi2016deep} proposed a more general privacy-preserving method DP-SGD, which adds differential privacy~\cite{dwork2006differential} for the stochastic gradient descent algorithm. Subsequently, some works~\cite{zhang2018differentially, pichapati2019adaclip, xu2020adaptive} focus on reducing the privacy cost of DP-SGD through adaptive clipping or adaptive learning rate. In addition, for membership inference attacks, some elaborate defense mechanisms, such as AdvReg~\cite{nasr2018machine} and MemGuard~\cite{jia2019memguard}, have been conceived to obscure the differences between the output posteriors of members and non-members.

\section{Conclusion} 

In this work, we introduce a new, more integrated membership signal: \textit{the Pattern of Metric Sequence}, which comes from various stages of model training.
We verify that this new signal not only includes these existing well-applied signals but also pays more attention to time-dependent patterns, such as fluctuations and correlations.
Based on this new signal, we propose a novel membership inference attack against ML models, named Sequential-metric based Membership Inference Attack.
We construct sequential versions of multiple metrics obtained from the training process of the target model (multi-metric sequences) and leverage an attention-based RNN to automatically mine the patterns of the metric sequences for inference.
Extensive experiments demonstrate \SeqMIA\ outperforms advanced baselines. 
we further conduct in-depth comparative analyses of metric non-sequences vs. metric sequences, and single vs. multiple metrics, revealing the reasons for its superior performance.
Then, we analyze some other factors on the attack performance. Additionally, we demonstrate that \SeqMIA\ outperforms existing advanced baseline attacks under several representative defenses. 
In the future, we aim to explore enhanced metrics with richer membership information and employ more efficient serialization models to further improve membership inference performance.

\section{Acknowledgements} 
This work is supported by National Key R\&D Program of China (2022YFB4501500, 2022YFB4501503).

\bibliographystyle{plain}
\bibliography{normal_generated}

\appendix

\section{Dataset Description}\label{app:dataset}
\mypara{CIFAR10/CIFAR100} CIFAR10 and CIFAR100~\cite{krizhevsky2009learning} are commonly used datasets for evaluating image recognition algorithms, each including 60,000 color images of size $32\times32$. The difference is only that the images in CIFAR10 are equally distributed into 10 classes, while the images in CIFAR100 are equally distributed into 100 classes.

\mypara{CINIC10} CINIC10~\cite{darlow2018cinic} contains 270,000 images within the same classes as CIFAR10. In particular, 60,000 samples belong to CIFAR10, while the other samples come from ImageNet~\cite{deng2009imagenet}.

\mypara{GTSRB} GTSRB~\cite{stallkamp2012man} is a benchmark dataset used for traffic sign recognition, which includes 51839 images in 43 classes. Since the size of these images is not uniform, we resize them to $32\times32$ pixels during data preprocessing.

\mypara{Purchase} Purchase is a dataset of shopping records with 197324 samples of 600 dimensions, which is extracted from Kaggle's ``acquire valued shopper'' challenge. Following previous works~\cite{shokri2017membership,salem2018ml,liu2022membership}, we cluster these data into 100 classes for evaluating membership inference attacks against non-image classifiers.

\mypara{News} News is a popular benchmark dataset for text classification. This dataset includes 20,000 newsgroup documents of 20 classes. Following~\cite{salem2018ml}, we convert each document into a vector of 134410 dimensions using TF-IDF.

\mypara{Location} Location is a preprocessed check-in dataset provided by Shokri et al.~\cite{shokri2017membership}, which is obtained from Foursquare dataset~\cite{yang2016participatory}. Location contains 5010 data samples of 446 dimensions across 30 classes.

\begin{table}[ht]
 \footnotesize
 \caption{Data splits for our evaluation.}
 \label{data-splits}
     \centering
     \scalebox{\tablescale}
     {
     \begin{tabular}{c | c c | c c | c}
         \toprule
          Dataset & $\mathcal{D}^t_{train}$ & $\mathcal{D}^t_{test}$ & $\mathcal{D}^s_{train}$ & $\mathcal{D}^s_{test}$ & $\mathcal{D}^k$\\
          \midrule 
          CIFAR10 & 10000 & 10000 & 10000 & 10000 & 20000\\
          CIFAR100 & 10000 & 10000 & 10000 & 10000 & 20000\\
          CINIC10 & 10000 & 10000 & 10000 & 10000 & 220000\\
          GTSRB & 1500 & 1500 & 1500 & 1500 & 45839\\
          Purchase & 20000 & 20000 & 20000 & 20000 & 110000\\
          News & 3000 & 3000 & 3000 & 3000 & 6000\\
          Location & 800 & 800 & 800 & 800 & 1400\\
          \bottomrule
     \end{tabular}
     }
\end{table}

\section{Additional Time-Dependent Patterns of Metric Sequences}\label{addtional-patterns}
\begin{figure}[h]
  \centering
  \includegraphics[width=\linewidth]{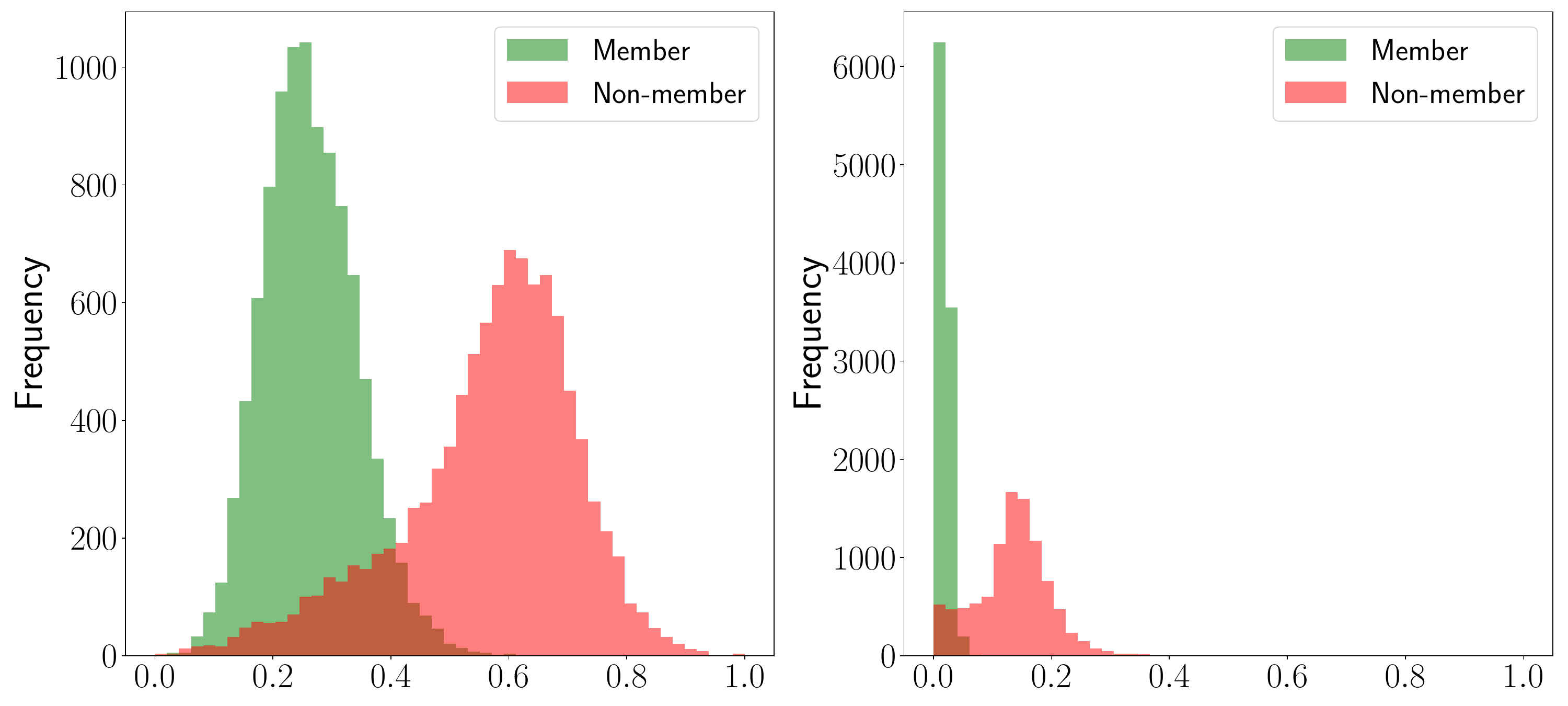}
  \caption{The distributions of the cumulative fluctuation amplitude of SD (left) and M-Entropy (right) within 100 epochs, which is obtained from 10,000 members and 10,000 non-members of VGG-16 trained on CIFAR100.}
  \label{appendix-metric-fluctuation}
\end{figure}

\begin{figure}[h]
  \centering
  \includegraphics[width=\linewidth]{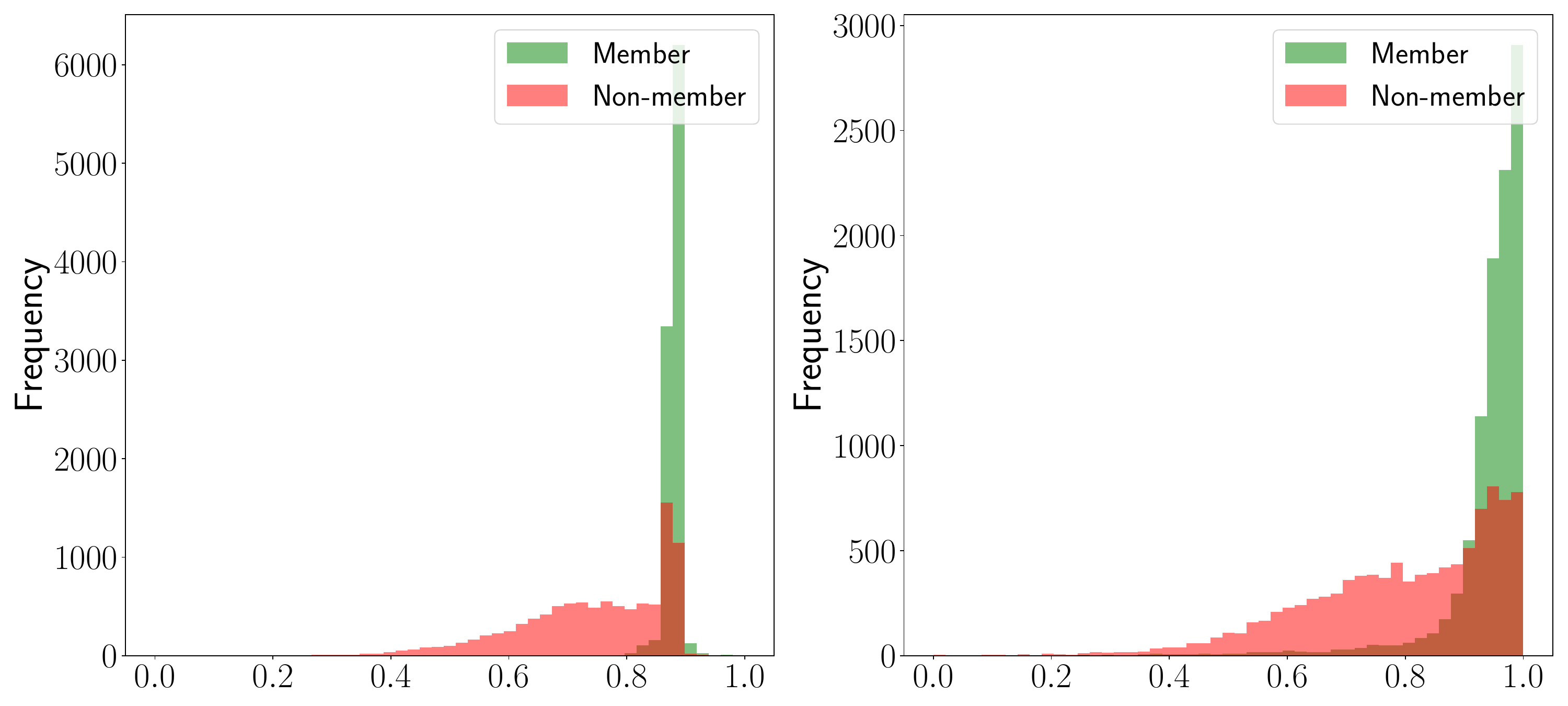}
  \caption{The distributions of the decline rate of Loss (left) and Entropy (right) within 100 epochs, which is obtained from 10,000 members and 10,000 non-members of VGG-16 trained on CIFAR100.}
  \label{appendix-metric-declinerate}
\end{figure}

\mypara{Decline Rate of Metric Sequences} We choose two metrics, Loss and Entropy, as our example. First, we construct the sequences of loss values and entropy values for each sample as the training progresses.
And the loss decline rate for each sample is calculated by measuring the loss decline amplitude within a period of \textit{consecutive epochs} and then dividing by the number of epochs. Similarly, we can obtain decline rate of entropy sequence. 
We then count the frequency of the samples regarding the distribution of their decline rate of loss (or entropy).
As depicted by \autoref{appendix-metric-declinerate}, we observe members exhibit significantly larger decline rate of loss (or entropy) sequence compared to non-members.
The results reconfirm that there exists a very clear difference in the pattern of metric sequence (e.g., loss sequence and entropy sequence) between members and non-members.

\section{Additional Experimental Results}

\begin{table*}[h]
\footnotesize
\caption{Attack performance of different attacks against ResNet-56 trained on four image datasets.}
\label{attack-performance-on-ResNet-56}
    \centering
    \scalebox{\tablescale}
    {
    \begin{tabular}{c c c c c c c c c c c c c}
        \toprule
         \multirow{2}{*}{MIA method} & \multicolumn{4}{c}{TPR @ 0.1\% FPR (\%)} &
         \multicolumn{4}{c}{Balanced accuracy} &
         \multicolumn{4}{c}{AUC}\\
         \cmidrule(r){2-5} \cmidrule(r){6-9} 
         \cmidrule(r){10-13} 
          & CIFAR10 & CIFAR100 & CINIC10 & GTSRB & CIFAR10 & CIFAR100 & CINIC10 & GTSRB & CIFAR10 & CIFAR100 & CINIC10 & GTSRB\\
         \midrule
         ST & 0.11 & 0.18 & 0.11 & 0.21 & 0.642 & 0.825 & 0.637 & 0.643 & 0.679 & 0.885 & 0.686 & 0.650\\
         MBA(Entropy) & 0.17 & 0.42 & 0.16 & 0.15 & 0.651 & 0.820 & 0.648 & 0.639 & 0.653 & 0.822 & 0.645 & 0.639\\
         MBA(M-Entropy) & 0.18 & 0.75 & 0.24 & 0.15 & 0.694 & 0.904 & 0.757 & 0.645 & 0.694 & 0.904 & 0.757 & 0.645\\
         LiRA & 0.30 & 0.06 & 0.00 & 0.00 & 0.577 & 0.719 & 0.617 & 0.503 & 0.601 & 0.767 & 0.632 & 0.388\\
         EnhancedMIA & 2.63 & 4.12 & 4.39 & 0.27 & 0.681 & 0.878 & 0.765 & 0.612 & 0.764 & 0.947 & 0.851 & 0.640\\
         TrajectoryMIA & 0.37 & 1.38 & 0.38 & 0.14 & 0.664 & 0.901 & 0.752 & 0.585 & 0.741 & 0.956 & 0.825 & 0.627\\
         \midrule
         SeqMIA & \textbf{4.43} & \textbf{21.67} & \textbf{6.89} & \textbf{0.62} & \textbf{0.723} & \textbf{0.925} & \textbf{0.791} & \textbf{0.666} & \textbf{0.813} & \textbf{0.975} & \textbf{0.883} & \textbf{0.731}\\
         \bottomrule
    \end{tabular}
    }
\end{table*}

\begin{table*}[!t]
\footnotesize
\caption{Attack performance of different attacks against WideResNet-32 trained on four image datasets.}
\label{attack-performance-on-WideResNet-32}
    \centering
    \scalebox{\tablescale}
    {
    \begin{tabular}{c c c c c c c c c c c c c}
        \toprule
         \multirow{2}{*}{MIA method} & \multicolumn{4}{c}{TPR @ 0.1\% FPR (\%)} &
         \multicolumn{4}{c}{Balanced accuracy} &
         \multicolumn{4}{c}{AUC}\\
         \cmidrule(r){2-5} \cmidrule(r){6-9} 
         \cmidrule(r){10-13} 
          & CIFAR10 & CIFAR100 & CINIC10 & GTSRB & CIFAR10 & CIFAR100 & CINIC10 & GTSRB & CIFAR10 & CIFAR100 & CINIC10 & GTSRB\\
         \midrule
         ST & 0.18 & 0.63 & 0.26 & 0.14 & 0.661 & 0.799 & 0.649 & 0.606 & 0.723 & 0.880 & 0.716 & \textbf{0.677}\\
         MBA(Entropy) & 0.19 & 0.43 & 0.23 & 0.15 & 0.679 & 0.809 & 0.675 & 0.639 & 0.686 & 0.818 & 0.697 & 0.643\\
         MBA(M-Entropy) & 0.19 & 0.44 & 0.27 & 0.15 & 0.703 & 0.844 & 0.752 & \textbf{0.641} & 0.703 & 0.844 & 0.752 & 0.641\\
         LiRA & 3.10 & 7.30 & 4.39 & 0.00 & 0.688 & 0.812 & 0.724 & 0.500 & 0.746 & 0.888 & 0.785 & 0.312\\
         EnhancedMIA & 0.15 & 0.12 & 0.16 & 0.10 & 0.591 & 0.588 & 0.652 & 0.536 & 0.627 & 0.593 & 0.679 & 0.529\\
         TrajectoryMIA & 0.36 & 2.35 & 0.67 & 0.55 & 0.648 & 0.861 & 0.767 & 0.565 & 0.728 & 0.930 & 0.861 & 0.630\\
         \midrule
         SeqMIA & \textbf{11.46} & \textbf{29.07} & \textbf{26.94} & \textbf{1.03} & \textbf{0.740} & \textbf{0.905} & \textbf{0.816} & 0.608 & \textbf{0.842} & \textbf{0.970} & \textbf{0.912} & 0.651\\
         \bottomrule
    \end{tabular}
    }
\end{table*}

\begin{table*}[!h]
\footnotesize
\caption{Attack performance of different attacks against MobileNetV2 trained on four image datasets.}
\label{attack-performance-on-MobileNetV2}
    \centering
    \scalebox{\tablescale}
    {
    \begin{tabular}{c c c c c c c c c c c c c}
        \toprule
         \multirow{2}{*}{MIA method} & \multicolumn{4}{c}{TPR @ 0.1\% FPR (\%)} &
         \multicolumn{4}{c}{Balanced accuracy} &
         \multicolumn{4}{c}{AUC}\\
         \cmidrule(r){2-5} \cmidrule(r){6-9} 
         \cmidrule(r){10-13} 
          & CIFAR10 & CIFAR100 & CINIC10 & GTSRB & CIFAR10 & CIFAR100 & CINIC10 & GTSRB & CIFAR10 & CIFAR100 & CINIC10 & GTSRB\\
         \midrule
         ST & 0.11 & 0.17 & 0.17 & 0.14 & 0.655 & 0.822 & 0.677 & 0.622 & 0.690 & 0.920 & 0.747 & 0.673\\
         MBA(Entropy) & 0.17 & 0.34 & 0.21 & 0.16 & 0.663 & 0.831 & 0.692 & 0.617 & 0.660 & 0.825 & 0.688 & 0.616\\
         MBA(M-Entropy) & 0.19 & 0.44 & 0.27 & 0.15 & 0.700 & 0.879 & 0.772 & 0.617 & 0.700 & 0.878 & 0.772 & 0.617\\
         LiRA & 0.55 & 0.03 & 0.56 & 0.07 & 0.573 & 0.752 & 0.603 & 0.501 & 0.589 & 0.804 & 0.635 & 0.414\\
         EnhancedMIA & 2.31 & 4.26 & 6.46 & 0.68 & 0.682 & 0.884 & 0.768 & 0.587 & 0.761 & 0.952 & 0.856 & 0.616\\
         TrajectoryMIA & 0.18 & 0.88 & 0.47 & 0.41 & 0.649 & 0.861 & 0.747 & 0.579 & 0.727 & 0.910 & 0.820 & 0.614\\
         \midrule
         SeqMIA & \textbf{5.55} & \textbf{24.26} & \textbf{13.00} & \textbf{1.16} & \textbf{0.724} & \textbf{0.922} & \textbf{0.806} & \textbf{0.648} & \textbf{0.815} & \textbf{0.979} & \textbf{0.899} &\textbf{0.709}\\
         \bottomrule
    \end{tabular}}
\end{table*}

\begin{table*}[!h]
\footnotesize
    \caption{Attack performance of different attacks against MLPs trained on three non-image datasets.}
\label{attack-performance-on-MLPs}
    \centering
    \scalebox{\tablescale}
    {
    \begin{tabular}{c c c c c c c c c c}
        \toprule
         \multirow{2}{*}{MIA method} & \multicolumn{3}{c}{TPR @ 0.1\% FPR (\%)} &
         \multicolumn{3}{c}{Balanced accuracy} &
         \multicolumn{3}{c}{AUC}\\
         \cmidrule(r){2-4} \cmidrule(r){5-7} 
         \cmidrule(r){8-10} 
          & News & Purchase & Location & News & Purchase & Location & News & Purchase & Location\\
         \midrule
         ST & 0.47 & 0.16 & 0.26 & 0.831 & 0.856 & 0.885 & 0.941 & 0.856 & 0.957\\
         MBA(Entropy) & 1.01 & 0.43 & 1.46 & 0.930 & 0.872 & 0.945 & 0.930 & 0.872 & 0.945\\
         MBA(M-Entropy) & 1.15 & 0.45 & 1.48 & 0.936 & \textbf{0.879} & 0.944 & 0.936 & 0.879 & 0.944\\
         LiRA & 4.84 & 0.09 & 1.18 & 0.795 & 0.549 & 0.699 & 0.847 & 0.562 & 0.757\\
         EnhancedMIA & 0.35 & 0.31 & 0.00 & 0.794 & 0.666 & 0.852 & 0.836 & 0.706 & 0.878\\
         TrajectoryMIA & 9.88 & 0.39 & 1.71 & 0.934 & 0.748 & 0.945 & 0.982 & 0.839 & 0.979\\
         \midrule
         SeqMIA & \textbf{31.16} & \textbf{6.39} & \textbf{25.23} & \textbf{0.938} & 0.876 & \textbf{0.969} & \textbf{0.989} & \textbf{0.938} & \textbf{0.992}\\
         \bottomrule
    \end{tabular}
    }
\end{table*}

\begin{table*}[ht]
\footnotesize
    \caption{Balanced accuracy and AUC of RNN-based SeqMIA, Transformer-based SeqMIA, and TrajectoryMIA against ResNet-56 trained on four image datasets.}
\label{acc-auc-rnn-vs-transformer}
    \centering
    \scalebox{\tablescale}
    {
    \begin{tabular}{c c c c c c c}
        \toprule
         \multirow{2}{*}{Dataset} & \multicolumn{3}{c}{Balanced accuracy} & \multicolumn{3}{c}{AUC}\\
         \cmidrule(r){2-4} \cmidrule(r){5-7} 
         & RNN & Transformer & TrajectoryMIA & RNN & Transformer & TrajectoryMIA\\
         \midrule
         CIFAR10 & \textbf{0.723} & 0.707 & 0.664 & \textbf{0.813} & 0.796 & 0.741\\
         CIFAR100 & \textbf{0.925} & 0.887 & 0.901 & \textbf{0.975} & 0.884 & 0.956\\
         CINIC10 & \textbf{0.791} & 0.723 & 0.752 & \textbf{0.883} & 0.854 & 0.825\\
         GTSRB & \textbf{0.666} & 0.539 & 0.585 & \textbf{0.731} & 0.572 & 0.627\\
         \bottomrule
    \end{tabular}
    }
\end{table*}

\begin{table*}[!h]
\footnotesize
\caption{Balanced accuracy and AUC for attacks using different membership signals against VGG-16 trained on four image datasets.}
\label{BA-and-AUC-ablation-study-on-VGG-16}
    \centering
    \scalebox{\tablescale}
    {
    \begin{tabular}{c c c c c c c c c}
        \toprule
         \multirow{2}{*}{MIA method} & \multicolumn{4}{c}{Balanced accuracy} & \multicolumn{4}{c}{AUC}\\
         \cmidrule(r){2-5} \cmidrule(r){6-9}
          & CIFAR10 & CIFAR100 & CINIC10 & GTSRB & CIFAR10 & CIFAR100 & CINIC10 & GTSRB\\
         \midrule
         Loss set & 0.666 & 0.892 & 0.730 & 0.540 & 0.739 & 0.949 & 0.811 & \textbf{0.666}\\
         Multi-metric set & 0.713 & 0.925 & 0.767 & 0.532 & 0.804 & 0.967 & 0.862 & 0.557\\
         \midrule
         Loss sequence & 0.719 & 0.940 & 0.805 & 0.558 & 0.824 & 0.986 & 0.905 & 0.594\\
         Multi-metric sequence & \textbf{0.766} & \textbf{0.959} & \textbf{0.850} & \textbf{0.577} & \textbf{0.869} & \textbf{0.992} & \textbf{0.937} & 0.649\\
         \bottomrule
    \end{tabular}
    }
\end{table*}

\begin{table*}[!h]
\footnotesize
    \caption{Performance for attacks using different membership signals against MLPs trained on three non-image datasets.}
\label{BA-and-AUC-ablation-study-on-MLPs}
    \centering
    \tabcolsep 5pt
    \scalebox{\tablescale}
    {
    \begin{tabular}{c c c c c c c c c c}
        \toprule
         \multirow{2}{*}{MIA method} &
         \multicolumn{3}{c}{TPR @ 0.1\% FPR (\%)} & \multicolumn{3}{c}{Balanced accuracy} & \multicolumn{3}{c}{AUC}\\
         \cmidrule(r){2-4} \cmidrule(r){5-7} \cmidrule(r){8-10}
          & News & Purchase & Location &
           News & Purchase & Location & News & Purchase & Location\\
         \midrule
         Loss set & 9.88 & 0.39 & 1.71 & 0.934 & 0.748 & 0.945 & 0.982 & 0.839 & 0.979\\
         Multi-metric set & 15.53 & 0.97 & 9.33 & 0.919 & 0.845 & 0.955 & 0.979 & 0.910 & 0.987\\
         \midrule
         Loss sequence & 22.52 & 5.51 & \textbf{26.54} & 0.924 & 0.832 & 0.949 & 0.981 & 0.911 & 0.987\\
         Multi-metric sequence & \textbf{31.16} & \textbf{6.39} & 25.23 & \textbf{0.938} & \textbf{0.876} & \textbf{0.969} & \textbf{0.989} & \textbf{0.938} & \textbf{0.992}\\
         \bottomrule
    \end{tabular}
    }
\end{table*}

\begin{table*}[h!]
\footnotesize
\caption{Balanced accuracy and AUC of SeqMIA against VGG-16 trained on CIFAR10 with DP-SGD.}
\label{BA-and-AUC-under-dpsgd}
    \centering
    \scalebox{\tablescale}
    {
    \begin{tabular}{c c c c c c}
        \toprule
          & \multicolumn{2}{c}{$\delta=1\textnormal{e-}5$ and $C=1$} &
         \multirow{2}{2cm}{Accuracy of the target model} &
         \multicolumn{2}{c}{Attack performance of SeqMIA}\\
         \cmidrule(r){2-3} \cmidrule(r){5-6} 
          & $\sigma$ & $\epsilon$ &  & Balanced accuracy & AUC\\
         \midrule
         No defense & - & - & 0.756 & 0.766 & 0.869\\
         \midrule
         \multirow{4}{*}{DP-SGD} & 0 & $\infty$ & 0.5746 & 0.676 & 0.733\\
          & 0.2 & 1523 & 0.581 & 0.518 & 0.524\\
          & 0.5 & 43 & 0.482 & 0.501 & 0.493\\
          & 1 & 6 & 0.3766 & 0.500 & 0.492\\
         \bottomrule
    \end{tabular}
    }
\end{table*}

\begin{table*}[!h]
\footnotesize
    \caption{Balanced accuracy and AUC of different attacks against ResNet-56 trained on CINIC10 with AdvReg and MixupMMD.}
\label{BA-and-AUC-against-defense}
    \centering
    \scalebox{\tablescale}
    {
    \begin{tabular}{c c c c c c c}
        \toprule
         \multirow{2}{*}{} & \multicolumn{3}{c}{Balanced accuracy} & \multicolumn{3}{c}{AUC}\\
         \cmidrule(r){2-4} \cmidrule(r){5-7} 
         & no defense & AdvReg & MixupMMD & no defense & AdvReg & MixupMMD\\
         \midrule
         ST & 0.637 & 0.516 & 0.518 & 0.686 & 0.922 & 0.687\\
         MBA(Entropy) & 0.648 & 0.500 & 0.503 & 0.645 & 0.500 & 0.503\\
         MBA(M-Entropy) & 0.757 & 0.754 & 0.567 & 0.757 & 0.751 & 0.565\\
         LiRA & 0.617 & 0.504 & 0.504 & 0.632 & 0.493 & 0.498\\
         EnhancedMIA & 0.765 & 0.686 & 0.560 & 0.851 & 0.746 & 0.581\\
         TrajectoryMIA & 0.752 & \textbf{0.834} & \textbf{0.632} & 0.825 & 0.910 & 0.733\\
         \midrule
         SeqMIA & \textbf{0.791} & 0.755 & 0.574 & \textbf{0.883} & \textbf{0.963} & \textbf{0.844}\\
         \bottomrule
    \end{tabular}
    }
\end{table*}

\begin{figure*}[h]
  \centering
  \includegraphics[width=0.85\linewidth]{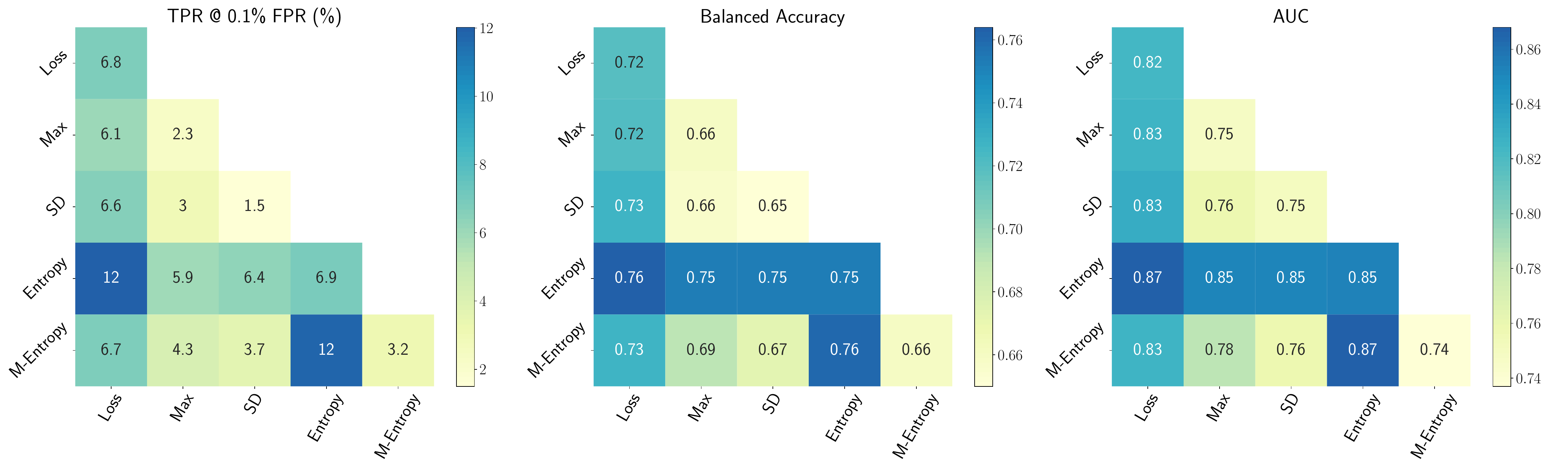}
  \caption{Attack performance of SeqMIA using double metrics against VGG-16 trained on CIFAR10.}
  \label{VGG-16-CIFAR10-double-metrics}
\end{figure*}

\end{document}